\documentclass[a4paper,11pt]{article}
\usepackage{jheppub}
\usepackage[T1]{fontenc}
\usepackage{latexsym}
\usepackage[latin1]{inputenc}
\usepackage{amsmath,epsfig}
\usepackage{amssymb,amsfonts}
\usepackage{bm}
\usepackage{mathrsfs}
\usepackage{cancel,empheq,dsfont,overpic,subcaption,subeqnarray}
\usepackage[normalem]{ulem}
\usepackage{slashed, braket}
\usepackage{xcolor}
\usepackage{color}
\let\normalcolor\relax
\usepackage{graphicx}
\usepackage{float, soul, parskip, lmodern}
\usepackage{caption}
\usepackage{siunitx}
\usepackage{subcaption}
\usepackage{amsmath}
\usepackage{longtable}
\usepackage{color}
\usepackage{bbm}
\usepackage{tikz}
\usepackage[compat=1.1.0]{tikz-feynman}
\usepackage{comment}

\def\hri#1#2{\href{http://arxiv.org/abs/#1}{[ArXiv:#1]#2}}
\def\hre#1#2{\href{http://arxiv.org/abs/#1/#2}{[ArXiv:#1/#2]}}

\def\hrj#1#2{\href{https://www.doi.org/#1}{#2}}
\relax
\renewcommand{\theequation}{\arabic{section}.\arabic{equation}}

\def\be{\begin{equation}}
\def\ee{\end{equation}}

\def\sla{\raise.15ex\hbox{$/$}\kern-.57em}

\newcommand{\bear}{\begin{eqnarray}}
\newcommand{\bea}{\begin{eqnarray}}
\newcommand{\bi}{\begin{itemize}}
\newcommand{\ba}{\begin{array}}
\newcommand{\eear}{\end{eqnarray}}
\newcommand{\eea}{\end{eqnarray}}
\newcommand{\ei}{\end{itemize}}
\newcommand{\ea}{\end{array}}
\def\hre#1#2{\href{http://arxiv.org/abs/#1/#2}{[ArXiv:#1/#2]}}

\newbox\pippobox

\def\a{\alpha}		\def\b{\beta}		\def\g{\gamma}		\def\d{\delta}
\def\e{\varepsilon}	\def\z{\zeta}					
\def\i{\iota}						\def\m{\mu}
\def\n{\nu}			\def\x{\xi}			\def\p{\pi}			\def\r{\rho}	
\def\s{\sigma}		\def\t{\tau}

\def\G{\Gamma}		\def\D{\Delta}

\def\<{\big\langle}
\def\>{\big\rangle}

\def\nn{\nonumber}
\newcommand{\black}[1]{{\color{black} #1 \color{black}}}

\def\sp{\;\;\;,\;\;\;}

\title{\boldmath
Anomalous and axial $Z'$ contributions to  $g-2$}

\preprint{CCTP-2022-2\\
\rightline{ITCP-2022/5}}

\author[a]{Pascal~Anastasopoulos}
\author[b]{Kunio~Kaneta}
\author[c]{Elias~Kiritsis}
\author[d]{Yann~Mambrini}

\affiliation[a]{\href{https://mathphys.univie.ac.at}{Mathematical Physics Group, Department of Physics, University of Vienna}, \\
~~ Boltzmanngasse 5, 1090 Vienna, Austria.}

\affiliation[b]{
\href{https://www.twcu.ac.jp/main/english/index.html}{Department of Mathematics, Tokyo Woman's Christian University},\\
~~ Tokyo 167-8585, Japan}

\affiliation[c]{\href{http://www.apc.univ-paris7.fr}{Universit\'e Paris Cit\' e, CNRS, Astroparticule et Cosmologie}, F-75013 Paris, France\\
 \centerline{\cal \&}\\
 ~~ \href{http://hep.physics.uoc.gr}{Crete Center for Theoretical Physics}, 
Department of Physics, \\
~~ University of Crete, 70013 Heraklion, Greece.}

\affiliation[d]{
\href{http://www.th.u-psud.fr}{Universit\'e Paris-Saclay}, CNRS/IN2P3, IJCLab, 91405 Orsay, France.}


\abstract{We study the effects of an anomalous $Z'$ boson on the
anomalous magnetic moment of the muon $(g-2)$, and especially  the impact of its axial coupling. We mainly evaluate the {\it negative} contribution to $(g-2)$ of such couplings at one-loop
and look at the anomalous couplings generated at two loops. We find areas of the parameter space, where the anomalous contribution becomes
comparable and even dominant compared to the one-loop contribution. We show that in such cases, the cutoff of the theory is sufficiently low, so that new charged fermions can be found in the next round of collider experiments.
We comment on the realization of such a context in string theory orientifolds.
}

\begin{document}
\maketitle

\section{Introduction}

The Standard Model (SM) is the most accurate physical theory ever made. It predicts, among others, the anomalous magnetic moment of the muon to be \cite{Aoyama:2020ynm} (see also \cite{other SM g-2 1, other SM g-2 2, other SM g-2 3, other SM g-2 4, other SM g-2 5, other SM g-2 6, other SM g-2 7, other SM g-2 8, other SM g-2 9, other SM g-2 10, other SM g-2 11, other SM g-2 12, other SM g-2 13, other SM g-2 14, other SM g-2 15, other SM g-2 16, other SM g-2 17, other SM g-2 18, other SM g-2 19, other SM g-2 20, other SM g-2 21, other SM g-2 22, other SM g-2 23, Crivellin:2020zul, other SM g-2 24, other SM g-2 25})
\begin{equation}
	 a^{\mathrm{SM}}_{\mu}=\SI{116591810(43) e-11}{}~.
	 \label{eq:AMMth}
\end{equation}
In comparison, the latest experimental measured value
of the Fermilab National Accelerator Laboratory (FNAL) Muon $g-2$ Experiment (E$989$) \cite{Abi:2021gix}
\footnote{See also \cite{Grange:2015fou,Bennett:2006fi} for BNL results and \cite{Abe:2019thb} for a planned experiment at J-PARC.}
\begin{equation}
	a^{\mathrm{exp}}_{\mu}(\mathrm{FNAL})=\SI{116592061(41)e-11}{}~,
	\label{eq:AMMexp}
\end{equation}
with a precision of $(0.35\; \mathrm{ppm})$\footnote{The interim result is analysed from the Run-$1$ dataset of $2018$. Later runs are being evaluated during the writing of this paper, and Run-$6$ is being planned. Furthermore, the E$24$ experiment at J-PARC will start in $2024$, which promises further accuracy \cite{Sako:2014fha}.}, gives
\begin{equation}
a_{\mu}^{\mathrm{exp}}-a_{\mu}^{\mathrm{SM}}= \SI{251(59)e-11}{}~.
\label{Eq:exp}
\end{equation}
The discrepancy is roughly of magnitude  $4.2\s$ and although not fully incompatible with the SM, it indicates the possibility  of  ``new physics'' contributions
\cite{Jackiw:1972jz,Bardeen:1972vi,Leveille:1977rc,Holdom:1985ag,Kukhto:1992qv,AK,Armillis:2008bg,Pospelov:2008zw,Freitas:2014pua,Davoudiasl:2014kua,Dorsner:2016wpm,1511.07447,1712.09360,Kamada:2018zxi,Crivellin:2018qmi,Liu:2018xkx,Dutta:2018fge,Han:2018znu,1906.11297,CarcamoHernandez:2019ydc,Jho:2019cxq,Endo:2019bcj,Badziak:2019gaf,Hiller:2019mou,Bauer:2019gfk,Gardner:2019mcl,Cornella:2019uxs,Crivellin:2019mvj,Abdullah:2019ofw,Criado:2019tzk,Dutta:2020scq,Yang:2020bmh,Bigaran:2020jil,Dorsner:2020aaz,Botella:2020xzf,Jana:2020pxx,CarcamoHernandez:2020pxw,Haba:2020gkr,Calibbi:2020emz,Arbelaez:2020rbq,Chen:2020jvl,Hati:2020fzp,Jana:2020joi,Chen:2020tfr,Chun:2020uzw,Li:2020dbg,Banerjee:2020zvi,Antoniadis:2021mqz,Hernandez:2021tii,Capdevilla:2021kcf,Barman:2021xeq,Hammad:2021mpl,Chowdhury:2021tnm,Yu:2021suw,Athron:2021iuf,Cadeddu:2021dqx,Bodas:2021fsy,Coy:2021wfs,Li:2021koa,Li:2021lnz,Han:2021gfu,Cao:2021lmj,Botella:2022rte,Kriewald:2022erk,DAlise:2022ypp,Panda:2022kbn,Chowdhury:2022jde,Anastasopoulos:2022wob,Crivellin:2021rbq}.
\footnote{Note that a recent lattice study on the hadronic contribution appears to ameliorate the discrepancy \eqref{Eq:exp} \cite{other SM g-2 21}.}
Some works  tried to reconciliate the $g-2$
excess with dark matter constraints \cite{Arcadi:2021zdk}, $B-$anomalies
\cite{Calibbi:2021vcq} or the $M_W$ curiosity \cite{Arcadi:2022dmt}.
In searching for new physics, one possible and rather generic scenario involves the extension of the SM with an additional massive gauge boson $Z'$. This gauge boson may couple to all or some SM fields (fermions and the Higgs) and presents one of the most popular scenarios for physics beyond the SM at the LHC. Among all the
possible
extensions of the SM, the introduction
of an additional gauge group is also one of the most motivated. It can for instance serve as a mediator between the dark
sector and the visible world, in order  to explain the dark matter
production in the early Universe \cite{Arcadi:2013qia}
and appear in unified theories \cite{Mambrini:2015vna}.

In this paper, we study an extension of the SM with {\em a more general} type of $Z'$ bosons, which are {\it superficially anomalous}. With the term ``superficially anomalous'' we refer to the fact that the associated gauge symmetry is anomaly-free in the ultraviolet (UV) fundamental theory, but appears anomalous in the low-energy effective field theory, that is obtained after integrating out an anomalous set of heavier states \cite{ABDK}. Such gauge symmetries are typically associated with gauged baryon, lepton number, or some Peccei-Quinn symmetry. Therefore, they are crucial for realizing the SM global symmetries and the associated constraints on baryon and lepton number conservation.

Depending on the realization of the UV theory, the interpretation of the appearance of low-energy anomalous gauge symmetries can be different. In string theory realizations,
such gauge symmetries are broken by the St\"uckelberg mechanism tied to the Green-Schwarz-type cancellation of their anomalies, and the associated gauge fields become massive \cite{ABDK, Kiritsis:2003mc, au1, AKR, PA2,MambriniGS,CIK,PA,Anastasopoulos:2008jt,Armillis:2008vp,Accomando:2016sge,Dudas:2012pb}. Additional generalized Chern-Simons terms (GCS) appear in the low energy theory to cure the model of all types of mixed anomalies, \cite{ABDK,AKT}.
Such anomalous $U(1)$'s in the effective theory, can also be interpreted as a UV completion in the context of QFT \cite{ABDK, NC}. The $U(1)$ gauge boson obtains a mass from one or more Higgs scalars, whereas the Higgs mechanism gives masses to an anomalous subset of fermions.
In the limit where the anomalous subset of fermions and the physical Higgses become heavy,  the low-energy theory is that of anomalous $U(1)$'s, with the St\"uckelberg scalar being the phase of the Higgs \cite{ABDK}\footnote{This is similar to the EFT of the SM after integrating out the (heavy) top quark.}.

The difference between anomalous and non-anomalous $Z'$s lies in a set of special, anomaly-related three-gauge-boson couplings which are  not present in the non-anomalous case \cite{AK, Armillis:2008bg, CIK, ABDK, PA, Anastasopoulos:2008jt, Armillis:2008vp, Accomando:2016sge} and can have
observable consequences on low energy phenomenology
\cite{CIK, Anastasopoulos:2008jt, Dudas:2012pb}.
Note that such effective theories can appear anomalous or non-anomalous at different scales.
Therefore, studying the phenomenological consequences of such models is exciting and relevant.

Adding abelian factors that couple to some or all of the SM fields produces very popular extensions of the SM.
In this category, we have
lepton-flavor violating $Z'$, leptophilic $Z'$s (abelian factors that couple only to leptons),
$U(1)_{T3R}$ (an abelian field that couples to all right-handed fermions of the SM), etc.
The non-anomalous versions of these constructions have been extensively studied. It becomes then interesting to look at the phenomenological consequences of their anomalous versions.
It is also worth mentioning that studying anomalous $U(1)$s is very timely due to upcoming experiments.
The Forward Physics Facility (FPF) is planned to operate near the ATLAS interaction point during the LHC high-luminosity era \cite{FPF}. It will investigate long-lived particles that might escape the apparatus, and new abelian gauge fields have a prominent position in these searches.

The present paper aims to evaluate the contribution of a $Z'$ boson to the $g-2$ of the muon\footnote{
Previous works evaluated the contribution of {\it non-anomalous} vector fields to the $g-2$,
\cite{Leveille:1977rc, Pospelov:2008zw, Davoudiasl:2014kua, 1511.07447, 1712.09360, 1906.11297, CarcamoHernandez:2019ydc, Hammad:2021mpl, Chowdhury:2021tnm, Yu:2021suw, Athron:2021iuf, Cadeddu:2021dqx, Kriewald:2022erk, DAlise:2022ypp, Panda:2022kbn},
without including the anomalous coupling.} including the possibility of anomalous couplings \cite{AK, Armillis:2008bg}. In particular, we focus on the leading contribution of the anomalous three-gauge-boson coupling to the $g-2$, comparing it to the contribution evaluated in the non-anomalous case. The anomalous couplings being
generated at one-loop, the difference
in the $g-2$ appears at two-loop order.
We focus on the most straightforward class of models, in which we extend the SM with a single additional anomalous abelian $U(1)$ gauge symmetry, an associated axion, and generalized Chern-Simons terms necessary to cancel all the possible anomalies. After electroweak symmetry breaking, we obtain the SM fields plus an additional massive $Z'$
whose non-minimal couplings
are entirely determined by
the anomaly cancellation conditions\footnote{In the presence of more than one $Z'$s this is no longer the case \cite{ABDK}.}. In a sense,
any phenomenological constraints on the couplings
give direct information on the content of the UV
completion due to anomaly matching, and vice-versa.

\subsubsection*{Results and outlook}

After parametrizing the anomalous three-gauge-boson couplings, we evaluate their contribution to the $g-2$ of the muon, first presented in general terms without using any assumption for the values of the parameters of our model.
Then, we proceed to a phenomenological analysis, where we consider a low-mass $Z'$ model (with $m_{Z'}\ll m_{Z}$), and we show that the anomalous contribution can dominate the two-loop diagrams while lying within the error zone of the discrepancy \eqref{Eq:exp}.

Indeed, in the leading one-loop contribution of a $Z'$ to the $g-2$, the axial and vectorial parts compete
with a negative relative sign.
Consequently, regions in the parameter space exist where the one-loop contribution can be highly
suppressed, becoming subdominant compared to higher-order contributions. In such a case, the anomalous (two-loop) diagrams that depend on the anomaly become dominant.
The anomalous contributions can be considered a litmus test for higher energy physics that filters to low energy via the anomalies.

We then calculate the anomalous contribution to the $g-2$, performing numerical integrations for a specific range of the mass of the $m_{Z'}$ and the couplings and compare the anomalous contribution to the non-anomalous one.
\bi
\item At one-loop, because  the vectorial and axial couplings of  the muon to an anomalous $Z'$ can be arbitrary and are not constrained by anomaly cancellation, the result is less constrained than non-anomalous models.

\item The anomalous $U(1)$ has an extra anomalous cubic coupling which depends on the overall anomaly, a property of the theory in the UV regime.
Any chiral fermion of the fundamental theory charged under the anomalous abelian factors contributes to the anomalous coupling. Therefore, apart from the SM fermions, the heavier fermions that are integrated out and are charged under the anomalous $Z'$, can give  a very large value to this coupling, with significant phenomenological consequences.
This  contribution can be comparable to the one-loop effects.
We show that it is much more significant to the standard non-anomalous $Z'$ two-loop contributions.

\item We also show that in the interesting cases, the cutoff of the theory is in the tens of TeV range, implying that there may be charged fermions to be discovered in the next generation of colliders.

\ei

In this paper, we evaluate the anomalous contribution to the $g-2$ of fermions, and we provide generic formulas as functions of the mass, the couplings of the $Z'$, and the anomalies. Some numerical examples show that the axionic and GCS contributions are leading compared to the fermionic triangle-loop.

We then explore some phenomenologically interesting models where the anomalous $Z'$ couples to the muon and tau (and maybe to other heavy fermions) but not to the electron, to avoid experimental constraints on the couplings and masses of the anomalous $Z'$.

Finally, we investigate whether an anomalous leptonic abelian factor that couples to the muon and tau and not to the electron can be found in semi-realistic D-brane configurations.
We analyze models with three, four, five, and more D-branes stacks, and we conclude that this scenario can be observed only in the six D-brane stack models.

The paper is organized as follows. In Sec.~\ref{SM+U(1)+axion} we present our EFT  model, i.e., the SM extended with an anomalous $U(1)$ and an axion. We explain our methodology to compute the anomalous three-gauge boson coupling diagrams in Sec.~\ref{The tiangle diagram}
and evaluate analytically the contribution of the anomalous coupling to the $(g-2)$ of the muon in Sec.~\ref{g-2}.
We then lead a complete numerical
analysis on the corresponding parameter space in Sec.~\ref{Sec:ALowMassZ'ModelAndResults}.
Finally, Sec.~\ref{D-branes} is devoted to a discussion about D-brane realizations of the SM, which could accommodate anomalous $U(1)$s with the properties mentioned in the previous sections.

In the appendixes, we provide some of the necessary ingredients of our model and give some more details on our computations.

\section{Extending the Standard Model with a single anomalous $U(1)_A$ and an axion}\label{SM+U(1)+axion}

{In this section, we introduce the effective theory of an extension of the SM with a single (massive) $U(1)_A$ and an associated axion.
This theory can be obtained as an effective theory,  from a class of theories  described in appendix \ref{An UV completion} where first we describe the generic properties.  In 
In subsection \ref{A low mass $Z'$ (toy) model}, we provide values to obtain a low mass anomalous $Z'$ model, which will be used in the later phenomenological part of this work.
}

\subsection{Building the Lagrangian}

When one extends the Standard Model by an extra ``anomalous" $U(1)_A$ and an axion $a$.
The charges of the SM fields under this extra $U(1)_A$ are assumed to be {family-non-universal}.
The transformations of the SM fields under this extra gauge symmetry, as well as the hypercharge symmetry, are
\bea
\ba{llllllllll}
& A^\m &\to & A^\m +\partial^\m \e, & ~~~ & Y^\m &\to & Y^\m +\partial^\m \z, \\
& a &\to & a-M \e,&~~~&\\
& \psi &\to & e^{ig_A q^{\psi-A} \e} \psi, &~~~& \psi &\to & e^{ig_Y q^{\psi-Y} \z} \psi,\\
& H &\to & e^{ig_A q^{H-A} \e} H,&~~~&  H &\to& e^{ig_Y q^{H-Y} \z} H,
\label{Eq:transform}
\ea
\eea
where $M= g_A q^{S-A} V$ is the mass of the anomalous $U(1)_A$, $A^\m$, and $\psi$ denotes all SM fermions collectively. $H$ is the SM Higgs doublet\footnote{In orientifold vacua realizing the SM spectrum, one always obtains two distinct Higgs fields.}
of the SM and ($\e$,$\z$) are the $U(1)_A$ and $U(1)_Y$ transformation parameters, respectively.

The charges of the SM fields are written
\bea
\ba{llccccrrcccccc}
&&&SU(3)&&SU(2)&&U(1)_Y&&U(1)_A&&\\
&Q^i_L&~:~&{\bf 3}&&{\bf 2}&&1/6~~&&q_i^{Q-A}&&\\
&u_R^{i,c}&~:~&\bar{\bf 3}&&{\bf 1}&&-2/3~~&&q_i^{u-A}&& \\
&d_R^{i,c}&~:~&\bar{\bf 3}&&{\bf 1}&&1/3~~&&q_i^{d-A}&& \\
&L_L^i&~:~&{\bf 1}&&{\bf 2}&&-1/2~~&&q_i^{L-A}&& \\
&l_R^{i,c}&~:~&{\bf 1}&&{\bf 1}&& 1~~&&q_i^{l-A}&&  \\
&H&~:~&{\bf 1}&&{\bf 2}&&1/2~~&&q^{H-A}&&
\ea
\label{Q_Acharges}\eea
with $i=1,2,3$ for the three SM families, and we use the hypercharge definition $Y=Q-T^3_L$.
Assuming that the Yukawa terms are allowed by all symmetries, we obtain the constraints\footnote{We do not assume charge universality, for reasons that will be clear in the phenomenological part of this work.}
\bea
\ba{llllllll}
\bar Q_L^i u_R^j H^*&~~~&:&~~~&-q_i^{Q-A}-q_j^{u-A}-q^{H-A}=0 \\
\bar Q_L^i d_R^j H&&:&&-q_i^{Q-A}-q_j^{d-A}+q^{H-A}=0 \\
\bar L_L^i l_R^j H&&:&&-q_i^{L-A}-q_j^{l-A}+q^{H-A}=0
\ea
\eea
Therefore, only few charges remain free.
However, this extension might suffer from mixed anomalies. For the generic charges, the anomaly-related traces are given below.
\bea
\ba{llllllll}
Tr[q_Y]=Tr[q_Yq_Yq_Y]=Tr[q_Y T_{SU(2/3)}^a T_{SU(2/3)}^a] &= 0 \\
Tr[q_Aq_Yq_Y]=\sum_{i}\frac{q_i^{Q-A}}{6} + \frac{4q_i^{u-A}}{3}+\frac{q_i^{d-A}}{3} +\frac{q_i^{L-A}}{2}+{q^{l-A}_i} ~~&= t_{AYY}\\
Tr[q_Yq_Aq_A]=\sum_{i}(q_i^{Q-A})^2 - 2(q_i^{u-A})^2 + (q_i^{d-A})^2 - (q_i^{L-A})^2 + (q_i^{l-A})^2 ~~&= t_{YAA}\\
Tr[q_Aq_Aq_A]=\sum_{i} 6(q_i^{Q-A})^3 + 3(q_i^{u-A})^3 + 3(q_i^{d-A})^3 + 2 (q_i^{L-A})^3 + (q_i^{l-A})^3 &= t_{AAA}~~~\\
Tr[q_AT^a_{SU(2)}T^a_{SU(2)}]= \sum_{i}3 q_i^{Q-A} + q_i^{L-A} ~~~~~~~~~~~~~~~~~~~~~~~~~~~&= T_{A,2}\\
Tr[q_AT^a_{SU(3)}T^a_{SU(3)}]= \sum_{i}2q_i^{Q-A} + q_i^{u-A} + q_i^{d-A} ~~~~~~~~~~~~~~~~~~~~&= T_{A,3}
\ea
\label{Aanomalies}\eea
where the traces are over all the SM fermions circulating in the anomalous loop and $T^a_{SU(2/3)}$ the generators of the non-abelian gauge groups of the SM. In our conventions, the $T^a_{SU(2/3)}$ normalization is
$Tr[T^a_i T^b_j]=\d^{ab}\d_{ij}/2
$ where $ij$ run over the different gauge groups, and $ab$ over the different generators of the relevant gauge groups.
{As we have already mentioned, these non-vanishing anomalies in the effective action are canceled via axionic and GCS terms, which appeared after integrating out the heavy fermions.}
The Lagrangian can then be written
\bea
{\cal L} &=& {\cal L}_\textrm{SM} +  {\cal L}_\textrm{extra~fields}
\label{Ltotal}\eea
where ${\cal L}_\textrm{SM}$ is the SM Lagrangian
and
\bea
{\cal L}_\textrm{extra~fields} &=&
-{1\over 4} F^A_{\mu \nu} F_A^{\mu \nu}
+{1\over 2} (\partial_\mu a +M A_\mu)^2+  g_A q^{\psi-A} A_\mu\bar \psi \g^\m\psi
 \nn\\
&&+2 i g_A q^{H-A} A^\m H^\dagger \Big(\partial_\m - i g_2{T^\a W^\a_\m -i g_Y q^{H-Y} Y_\m - {i\over 2}  g_A q^{H-A} A_\m}\Big) H  \nn\\
&&+
{1\over 24\p^2}
a \Big( C_{YY} F_Y \wedge F_Y + C_{YA} ~F_Y \wedge F_A+ C_{AA} ~F_A \wedge F_A +\sum_{i=2,3} D_i ~Tr_i[G\wedge G] \Big) \nn\\
&&+
{1\over 24\p^2}
 A\wedge  Y \wedge \Big( E_{AY,A} F_A+ E_{AY,Y} F_Y \Big)\nn\\
&&+
{1\over 24\p^2}
\sum_{i=2,3}Z_{i}~ A\wedge
Tr_i\Big[A\wedge (d A-{2\over 3}A\wedge A\Big)\Big]
\label{Lextrafields}\eea
where
\bea
F_i\wedge F_j={1\over 4} \e_{\m\n\r\s} F_i^{\m\n}F_j^{\r\s}\sp A_i\wedge A_j\wedge F_k\equiv {1\over 2}\e_{\m\n\r\s}A_i^{\m}A_j^{\n}F_k^{\r\s}
\label{Eq:FF}
\eea
with $F_X^{\mu \nu}=\partial^\mu X^\nu-\partial^\nu X^\mu$
the field strength of the field $X_\mu$.
Also, by  $i=2,3$ we denote the non-abelian sectors of SU(2) and SU(3). $Tr_i$ is the appropriate non-abelian trace in the fundamental representation and $G$ stands for the non-abelian field strength.
Note that the two last lines in \eqref{Lextrafields} are the GCS terms for the abelian and abelian-non abelian factors, whereas GCS terms between the hypercharge $Y$ and the non-abelian factors vanish \cite{ABDK}.

In the second line of \eqref{Lextrafields} we have the minimal couplings of the anomalous $U(1)_A$ to the SM fields, fermions $\psi$ = quarks, leptons with charges $q^{\psi-A}$ and to the Higgs field $H$ with charge $q^{H-A}$ (see \eqref{Q_Acharges}). Notice the square $A^\m A_\m$ term from the Higgs coupling, which provides an extra contribution to the mass of the anomalous $U(1)_A$.
We should also mention the presence of instanton-generated couplings between the fermions and the axion, which we omit here, as they will not be relevant for our calculation.

At 1-loop, the gauge symmetries are broken due to the anomalous diagrams, i.e., the non-vanishing traces of \eqref{Aanomalies}. Therefore, the effective theory is not invariant due to the triangle contributions of the low-energy fermions that give
\bea \delta S_{\rm one-loop}&=&-
{1\over  24\p^2} \int
\bigg\{\epsilon\Big(t_{AYY} F_Y\wedge F_Y
+t_{YAA}F_A\wedge
F_Y+{t_{AAA}}F_A\wedge F_A
+T_{A,i} Tr[G_i\wedge G_i] \Big) \nn\\
&& ~~~~~~~~~~~ +\zeta \Big(t_{YAA}F_A\wedge F_A+t_{AYY}F_A\wedge F_Y\Big) \bigg\}\label{dS_1loop}\eea
where the coefficients are related to the low-energy charges via (\ref{Aanomalies}).
We shall assume that such coefficients are generically non-zero.

Because of this, not only the extra $U(1)_A$ is anomalous, but also hypercharge has an anomaly, as testified by the terms in \eqref{dS_1loop} that are proportional to $\z$.
Of course, these anomalies should be canceled if we also include the anomalous variation of the tree level effective action
 in (\ref{Lextrafields}).
Imposing anomaly cancellation gives (where $i=2,3$ for the non-abelian factors)
\bea
&&
\d S_\textrm{one-loop} + \d S_{\rm extra~ fields}=0
~ \to ~\left\{
\ba{llllllll}
MC_{YY}=-2 {g_Ag_Y^2} t_{AYY} &&MC_{YA}=-2{g_A^2g_Y} t_{YAA}\\
MC_{AA}= - {g_A^3} t_{AAA}
&&  2M D_i = - 3Z_{i} = -6 {g_A g_i^2} T_{A,i}\\
E_{AY,A}= {g_A^2g_Y}t_{YAA}&& E_{AY,Y}= {g_Ag_Y^2} t_{AYY}
\ea\right.~~~~~~~
\label{anomalyConditionAY}
\eea
The conditions \eqref{anomalyConditionAY} fix the coefficients of the axionic and the GCS terms in (\ref{Lextrafields}).
Note that when there are at least two St\"ukelberg gauge fields and an analogous number of axions in the action, an extra gauge-invariant term involving the generalized Chern-Simons terms is available, \cite{ABDK},
\bea
{\cal L}_{inv}={1\over 2} \e_{\m\n\r\s} (\partial^\m a_I +M_I A_I^\m)(\partial^\n a_J +M_J A_J^\n)({\mathbbm d}_4^{IJK} F^{\r\s}_{A_K}+{\mathbbm d}^{IJ}_{5} F^{\r\s}_Y)
\label{GIextraTERM}\eea
The ${\mathbbm d}_{4,5}$ are constants, which we call gauge-invariant GCS. These couplings are not determined by the anomalies (like the GCS).
The low energy coefficient for such a term cannot be obtained by anomaly considerations alone.
In our model (with just one St\"ukelberg field $A^\m$ and a single axion $a$), a term like \eqref{GIextraTERM} is not present due to antisymmetry.

\subsection{The mass spectrum}\label{Masses of Zs}

The electroweak symmetry breaking alters the situation significantly because the Higgs is also charged under the extra $U(1)_A$ gauge symmetry:
\bi
\item We have five real scalar fields putting together the electroweak Higgs (4) and the axion (1). The Higgs field mixes with the axion. We obtain
\bi
\item The actual Standard Model Higgs with a mass
$m \simeq 125$ GeV whereas
\item The other four combinations are Goldstone modes which will
become the longitudinal polarization of the four gauge bosons $W^\pm, Z,Z'$.
\ei
\item We have the breaking of $SU(2)\times U(1)_Y\times U(1)_A\to U(1)_{em}$.
\bi
\item $W^+,W^-$ are the same combinations of $W^{1,2}$ with the original SM case. They absorb two Goldstone modes and become massive.
\item The rest of the $Y_\m$, $A_\m$, and $W^3_\m$ are mixed. Upon diagonalization, they give a massless gauge field,  the photon $\gamma$, and two massive fields $Z,Z'$ (for a more detailed discussion, see section (4.1) in \cite{CIK}).
\ei
\ei
Expanding the Higgs and axionic sector of \eqref{Ltotal}
\bea
|D_\m H|^2 + {1\over 2} (\partial_\m a +M A_\m)^2
\eea
with
\bea
D_\m H= \scalebox{0.97}{$\left(\ba{cccccc}
\partial_\m +{i\over 2} g_2 W_{3\m} + {i\over 2} g_Y Y_{\m} + {i\over 2} g_A q^{H-A} A_{\m} &  {i\over \sqrt{2}} g_2 W^+_{\m}\\
{i\over \sqrt{2}} g_2 W^-_{\m} & \partial_\m -{i\over 2} g_2 W_{3\m} + {i\over 2} g_Y Y_{\m} + {i\over 2} g_A q^{H-A} A_{\m}
\ea\right) $} H  \nn\\
\eea
and assuming that the vev of the Higgs has the form $\langle H\rangle =
(0,v)^T$, the mass matrix of the neutral gauge bosons reads
\bea
M^2_V &=
{1\over 4}\left(
\ba{cccccc}
g_Y^2v^2 && -g_2 g_Y v^2 && g_Y vN
\\
-g_2g_Y v^2 && g_2^2 v^2 && -g_2 vN
\\
g_Y vN
&& -g_2 vN
&& 2M^2+{ N^2 }\\
\ea
\right).
\eea
where, $M= g_A q^{S-A} V$, $N=g_Aq^{H-A} v$ and the field basis is taken as $(Y, W_3, A)$ \cite{CIK,Buras}.
By a real orthogonal matrix
\bea
{\cal O} &=
\left(
\ba{cccccc}
\cos\theta_W^0 && \sin\theta_W^0 && 0 \\
-\cos\zeta_+\sin\theta_W^0 && \cos\zeta_+\cos\theta_W^0 && \sin\zeta_+ \\
\cos\zeta_-\sin\theta_W^0 && \cos\zeta_-\cos\theta_W^0 && \sin\zeta_-
\ea
\right),
\eea
the mass matrix can be diagonalized as
\bea
{\cal O}M^2_V{\cal O}^T &= {\rm diag}\left[0,\frac{m_-^2+2g^2v^2}{8},\frac{m_+^2+2g^2v^2}{8}\right],
\label{massesofGaugeBosons}\eea
where we defined
\bea
&&
g^2\equiv g_Y^2+g_2^2,~~~\sin\theta_W^0 \equiv g_Y/g,~~~\cos\theta_W^0\equiv g_2/g,\\
&&
m_\pm^2 \equiv 2M^2+N^2-g^2v^2\pm\sqrt{(2M^2+ N^2-g^2v^2)^2+4g^2v^2N^2},\\
&&
\sin\zeta_\pm \equiv \frac{2g v N}{\sqrt{m_\pm^4+4g^2v^2N^2}},~~~\cos\zeta_\pm\equiv \frac{m_\pm^2}{\sqrt{m_\pm^4+4g^2v^2N^2}}.
\eea
{Notice that for small coupling $q^{H-A}$, $N\to 0$ and we restore the pure SM couplings.}

In the  various limits we have
\bi
\item At $M\to\infty$, it is
\bea
&&m_+^2\simeq 4M^2+{\cal O}(1)\sp m_-^2\simeq -{g^2 v^2N^2\over M^2}+{\cal O}(M^{-4})\;,\\
&&\sin\zeta_+\simeq {gvN\over 2M^2}+{\cal O}(M^{-4})\sp
\cos\zeta_+\simeq 1+{\cal O}(M^{-4})\sp\\
&&\sin\zeta_-\simeq 1+{\cal O}(M^{-4})\sp
\cos\zeta_-\simeq -{g v N\over 2 M^2}+{\cal O}(M^{-4}).
\eea
We, therefore, recover the mass spectrum
\be
(m_-^2+2g^2v^2)/8\simeq g^2v^2/4-g^2v^2N^2/8M^2\sp (m_+^2+2g^2v^2)/8\simeq M^2/2\;.
\ee
\item On the other hand, in the limit $M\to 0$, we obtain instead
\bea
&&m_{+}^2\simeq
 2g_A^2(q^{H-A})^2 v^2+{4g_A^2(q^{H-A})^2 \over g_A^2(q^{H-A})^2 +g^2 } M^2+{\cal O}(M^4),\\
&&m_-^2\simeq -2g^2 v^2 +
{4g^2 \over  g_A^2(q^{H-A})^2 +g^2}M^2+{\cal O}(M^4),\\
&&\sin\zeta_+\simeq {g v N\over \sqrt{g^2v^2N^2+(N^2+g^2v^2)^2}}+{\cal O}(M^2),\\
&&\cos\zeta_-\simeq 
{2g\over g_Aq_A^H}{M^2\over N^2+g^2v^2}+{\cal O}(M^4).
\eea
Thus, the mass spectrum becomes
\bea
&&(m_-^2+2g^2v^2)/8\simeq {1\over 2}{g^2M^2\over g^2 +g_A^2 (q^{H-A})^2}
+{\cal O}(M^4),\\ &&(m_+^2+2g^2v^2)/8\simeq {1\over 4}(g^2+g_A^2 (q^{H-A})^2) v^2
+{\cal O}(M^4)\;.
\eea

\ei
For small $M$, we have an important constraint on the validity of the effective theory derived first in \cite{Preskill}.
The cutoff $\Lambda$ for the effective description must be smaller than
\be
\Lambda\lesssim 64\pi^3{M\over g_A(g_A^2 t_{AAA}+2g_Ag_Y t_{YAA}+g_Y^2 t_{AYY}+\sum_{i=2}^3g_i^2 T_{A,i})}
\label{thecutoff}
\ee
This must be estimated for the cases of interest, which is done later in subsection \ref{the anomalous contribution}.
As is clear, the cutoff lowers as the mass becomes smaller, or the traces become larger, or the gauge couplings become stronger.
Finally, the relations between gauge fields in the unbroken and the broken phase $(\g,Z,Z')$ are given by
\bea
Y &=&
\cos\theta_W^0 \g-\cos\zeta_+\sin\theta_W^0 Z+\cos\zeta_-\sin\theta_W^0 Z',\nn\\
W_3 &=&
\sin\theta_W^0 \g+\cos\theta_W^0\cos\zeta_+Z-\cos\theta_W^0\cos\zeta_-Z',\nn\\
A &=&
\sin\zeta_+Z-\sin\zeta_-Z'.
\label{YW3A to gamma Z Z' basis}\eea
and the couplings/charges in the two basis are
\bea
g_Y q^{f-Y} &=&
\cos\theta_W^0 g_\g q^{f-\g}-\cos\zeta_+\sin\theta_W^0 g_{Z} q^{f-Z} +\cos\zeta_-\sin\theta_W^0 g_{Z'}q^{f-Z'},\nn\\
g_2 q^{f-W_3} &=&
\sin\theta_W^0 g_\g q^{f-\gamma} +\cos\theta_W^0\cos\zeta_+g_{Z}q^{f-Z}-\cos\theta_W^0\cos\zeta_-g_{Z'}q^{f-Z'},\nn\\
g_A q^{f-A} &=&
\sin\zeta_+g_Zq^{f-Z}-\sin\zeta_-g_{Z'}q^{f-Z'}.
\label{gYgW3gA to ggamma gZ gZ' basis}\eea

Note that $A$ does not have a photon component, and therefore terms like $\g \wedge (Z/Z')\wedge F_{\gamma}$ may appear only from $A \wedge Y\wedge F_Y$.

\subsection{The generalized Chern-Simons terms in the mass-eigenstate basis}\label{GCS in mass eigenstate basis}


We can compute the effective couplings below the EWSB scale in the basis
of mass eigenstates. We obtained

\bea
A \wedge Y\wedge F_Y &=&c_{ZZ'Z}Z\wedge Z'\wedge F_Z+c_{ZZ'Z'}Z\wedge Z'\wedge F_{Z'}+c_{Z\g \g}Z\wedge A_{\gamma}\wedge F_{A_{\gamma}}\nn\\
&&+c_{Z'\g \g}Z'\wedge A_{\g}\wedge F_{A_{\g}}
+c_{\g ZZ}A_{\g}\wedge Z\wedge F_Z+c_{\g Z'Z'}A_{\g}\wedge Z'\wedge F_{Z'}\nn\\
&&+c_{\g Z'Z} A_{\g}\wedge Z'\wedge F_Z+c_{ZZ"\g}Z\wedge Z'\wedge F_{A_{\g}}
\eea
with\footnote{Where we have used
\be
\int Z\wedge \g\wedge F_{Z'}=-\int \g\wedge Z'\wedge F_Z+\int Z\wedge Z'\wedge F_{A_{\g}}
\ee
}
\bea
&&c_{ZZ' Z}=-\cos\zeta_+\sin(\zeta_++\zeta_-)\sin^2\theta^0_W
\sp
c_{ZZ'Z'}=\cos\zeta_-\sin(\zeta_++\zeta_-)\sin^2\theta^0_W~~~~~~~~\\
&&c_{Z\g \g}=\sin\zeta_+\cos^2\theta^0_W\sp c_{Z'\g\g}=\sin\zeta_-\cos^2\theta^0_W\\
&&c_{\g ZZ}={1\over 4}\sin(2\zeta_+)\sin(2\theta^0_W)\sp c_{\g Z'Z'}=-{1\over 4}\sin(2\zeta_-)\sin(2\theta^0_W)\\
&&c_{\g Z'Z}={1\over 2}\sin(\zeta_++\zeta_-)\sin(2\theta^0_W)\sp
c_{ZZ'\g}=-{1\over 2}\sin\zeta_-\cos\zeta_+\sin(2\theta^0_W)
\eea
and
\bea
A \wedge Y\wedge F_A &=&d_{ZZ'Z}Z\wedge Z'\wedge F_Z+d_{ZZ'Z'}Z\wedge Z'\wedge F_{Z'}+d_{Z\g \g}Z\wedge A_{\gamma}\wedge F_{A_{\gamma}}\nn\\
&&+d_{Z'\g \g}Z'\wedge A_{\g}\wedge F_{A_{\g}}
+d_{\g ZZ}A_{\g}\wedge Z\wedge F_Z+d_{\g Z'Z'}A_{\g}\wedge Z'\wedge F_{Z'}\nn\\
&&+d_{\g Z'Z} A_{\g}\wedge Z'\wedge F_Z
+d_{ZZ'\g}Z\wedge Z'\wedge F_{A_{\g}}
\eea
with
\bea
&&d_{ZZ' Z}=\sin\zeta_+\sin(\zeta_++\zeta_-)\sin\theta^0_W
\sp
d_{ZZ'Z'}=\sin\zeta_-\sin(\zeta_++\zeta_-)\sin\theta^0_W~~~~~~~~\\
&&d_{Z\g \g}=d_{Z'\g\g}=0\\
&&d_{\g ZZ}=-\sin^2(\zeta_+)\sin(\theta^0_W)\sp d_{\g Z'Z'}=-\sin^2(\zeta_-)\sin(\theta^0_W)\\
&&d_{\g Z'Z}=0\sp
d_{ZZ'\g}=\sin\zeta_-\sin\zeta_+\cos(\theta^0_W)
\eea

In particular, from (\ref{Lextrafields}), we can extract the two anomalous cubic vertices for the SM gauge bosons $\gamma$ and $Z$ as
\be
{\cal L}_{anom}^{cubic}={1\over 24\pi^2}\int\left[\xi_{Z\g\g} Z\wedge A_{\g}\wedge F_{A_{\g}}+\xi_{\g ZZ}A_{\g}\wedge Z\wedge F_Z\right]
\label{anp}\ee
with
\be
\xi_{Z\g\g}=E_{AY,Y} \times c_{Z\g\g}+E_{AY,A} \times d_{Z\g\g}
\sp
\xi_{\g ZZ}=E_{AY,Y} \times c_{\g ZZ}+E_{AY,A} \times d_{\g ZZ}
\label{GCS with no Z'}\ee

Finally, we present the cubic terms between the abelian factors $\g,Z,Z'$ coming from the abelian-non abelian GCS terms in \eqref{Lextrafields}. We will discuss separately mixed anomalies between the anomalous $A$ and $SU(2)$ and $SU(3)$
\bi
\item For the SU(2), we have $A^{a,\m}_2=W_a^\m$ with $a=1,2,3$ and the GCS term is
\bea
&&A\wedge
Tr\Big[W\wedge (d W+{2\over 3}W \wedge W\Big)\Big]
\eea
Using our normalization for the generators $Tr[T^aT^b]=\delta^{ab}/2$ and $Tr[T^aT^bT^c]=i/4$ for different $a,b,c$, otherwise the trace is zero, 
%
%
%
%
we find terms that mix $\g,Z,Z'$ with the $W^\pm$ bosons.
\bea
{\cal L}^{GCS}_{SU(2)}&=&{1\over 12\p^2}
T_{2}~ A\wedge
Tr\Big[W\wedge (d W+{2\over 3}W \wedge W\Big)\Big]\\
&=&
{1\over 24\p^2}
T_{2}~ \Big(\sin ( {\z_2}) ~ { F_{W^-} }  \wedge  { {W^+} }  \wedge Z'   -\sin (\z_1) ~ { F_{W^-} }  \wedge  { {W^+} }  \wedge Z \nn\\
&&~~~~~~~~~~~~ +\sin ( {\z_2})~ { F_{W^+} }  \wedge  { {W^-} }  \wedge Z'  -\sin (\z_1) ~ { F_{W^+} }  \wedge  { {W^-} }  \wedge Z \nn\\
&&~~~~~~~~~~~~ -\sin  \theta^0_W  \cos  \theta^0_W  \sin ( {\z_1}) \cos ( {\z_2}) ~ {F_\gamma }  \wedge Z  \wedge Z' \nn\\
&&~~~~~~~~~~~~ +\sin  \theta^0_W  \cos  \theta^0_W  \cos ( {\z_1}) \sin ( {\z_2})  ~{F_\gamma }  \wedge Z  \wedge Z' \nn\\
&&~~~~~~~~~~~~ +\sin ^2 \theta^0_W  \sin   ( {\z_2}) ~ {F_\gamma }  \wedge \gamma   \wedge Z'  +\sin ^2 \theta^0_W  \sin ( {\z_1}) ~ {F_\gamma }  \wedge Z  \wedge \gamma  \nn\\
&&~~~~~~~~~~~~ -\cos ^2 \theta^0_W  \sin ( {\z_1}) \cos ^2( {\z_2}) ~Z  \wedge  {F_{Z'}}  \wedge Z' \nn\\
&&~~~~~~~~~~~~ +\cos ^2 \theta^0_W  \cos ( {\z_1}) \sin ( {\z_2}) \cos ( {\z_2})   ~Z  \wedge  {F_{Z'}}  \wedge Z' \nn\\
&&~~~~~~~~~~~~ +\sin  \theta^0_W  \cos  \theta^0_W  \sin ( {\z_2}) \cos ( {\z_2}) ~\gamma   \wedge  {F_{Z'}}  \wedge Z' \nn\\
&&~~~~~~~~~~~~ -\sin  \theta^0_W  \cos  \theta^0_W  \sin   ( {\z_1}) \cos ( {\z_2}) ~Z  \wedge \gamma   \wedge  {F_{Z'}} \nn\\
&&~~~~~~~~~~~~ +\sin  \theta^0_W  \cos  \theta^0_W  \cos ( {\z_1}) \sin ( {\z_2})  ~{F_{Z}}  \wedge \gamma  \wedge Z' \nn\\
&&~~~~~~~~~~~~ -\cos ^2 \theta^0_W  \sin ( {\z_1}) \cos ( {\z_1}) \cos ( {\z_2}) ~ {F_{Z}}  \wedge Z  \wedge Z' \nn\\
&&~~~~~~~~~~~~ +\cos ^2 \theta^0_W  \cos ^2( {\z_1}) \sin ( {\z_2})~{F_{Z}}  \wedge Z  \wedge Z' \nn\\
&&~~~~~~~~~~~~ +\sin  \theta^0_W  \cos  \theta^0_W  \sin ( {\z_1}) \cos ( {\z_1}) ~ {F_{Z}}  \wedge Z  \wedge \gamma \nn\\
&&~~~~~~~~~~~~ +\cos  \theta^0_W  \sin ( {\z_1}- {\z_2}) ~ {W^-} \wedge  {W^+} \wedge Z \wedge Z' \nn\\
&&~~~~~~~~~~~~
-\sin  \theta^0_W  \sin ( {\z_2}) ~ {W^-} \wedge  {W^+} \wedge \gamma \wedge Z' \nn\\
&&~~~~~~~~~~~~
-\sin  \theta^0_W  \sin ( {\z_1}) ~ {W^-} \wedge  {W^+} \wedge Z \wedge \gamma
\Big)
\nn\eea
\item For SU(3), we have $A^{a,\m}_3=G^\m_a$ for $a=8$ and the GCS terms mix $\g,Z,Z'$ bosons with the CS term of the gluons
\bea
{\cal L}^{GCS}_{SU(3)}&=&{1\over 12\p^2}
T_{3}~ A\wedge
Tr\Big[G\wedge (d G+{2\over 3}G \wedge G\Big)\Big]\\
&=&
{1\over 12\p^2}
T_{3}~ \bigg( \sin (\z_1) Z - \sin (\z_2) Z'
\bigg)\wedge
Tr\Big[G\wedge (d G+{2\over 3}G \wedge G\Big)\Big]\nn
\eea
which for non vanishing mixed anomaly $T_3$ we have couplings of the $Z,Z'$ to the gluons.
\ei

{That result is expected when there are mixed anomalies between $Y^\m$ and $A^\m$. Going to the photon basis, such anomalies will appear as anomalies between the photon and $Z$ and $Z'$, which can only be canceled via GCS terms of the form \eqref{anp} since no axionic terms are allowed for the photon (they would render the photon massive).}

Using the results above, we find that in the limit $M\to\infty$, both $\xi_{Z\g\g}$ and $\xi_{\g ZZ}$ vanish as $1/M^2$, consistent with decoupling.
On the other hand, in the opposite limit, $M\to 0$ both $\xi_{Z\g\g}$ and $\xi_{\g ZZ}$ reach finite non-zero limits.

{
The triple gauge boson coupling such as $\xi_{Z\gamma\gamma}$ and $\xi_{\gamma ZZ}$ may be constrained by collider experiments. The constraints are typically given to the triple gauge boson operators written in the $U(1)_{em}$ invariant form, namely, apparent $A_\gamma$ is replaced with $F_{A_\gamma}$ in the operators, which is, however, not exactly the case of the GCS terms \cite{Hagiwara:1986vm,Barklow:1996in}.
Nevertheless, conservatively we estimate $\xi_{\gamma ZZ}$, $\xi_{Z\gamma\gamma}\lesssim 10^{-2}$ by incorporating the loop factor \cite{Biekotter:2021int}, though one should be careful in using this limit since different momentum dependence than the GCS terms in the triple gauge boson coupling is assumed.
}

\subsection{Fixing the Gauge}

Expanding the Lagrangian in the broken phase \eqref{Ltotal}, we obtain terms of the form
\bea
C_i X_\m \partial^\m {\cal G}_i
\label{terms to be removed}
\eea
where $X_\m\equiv Z, Z', W^\pm$. $C_i$ are constants with the dimension of masse, and ${\cal G}_i$ are the 4 Goldstone bosons (three from the Higgs and one from the axion).

The terms in \eqref{terms to be removed} can be removed in the $\xi$-gauge
\bea
{\cal L}_{gauge~fixing} &=&-{1\over 2 \xi_\g} \Big(\partial_\m A_\g^\m)^2\label{gaugefixingALL}\\
&&-{1\over 2 \xi_Z} \Big(\partial_\m Z^\m - \xi_Z \sum_{i} C_{Z,i} {\cal G}_i \Big)^2\nn\\
&&-{1\over 2 \xi_{Z'}} \Big(\partial_\m {Z'}^\m - \xi_{Z'} \sum_{i} C_{{Z'},i} {\cal G}_i \Big)^2\nn\\
&&-{1\over \sqrt{\xi_+\xi_-}}
\Big(\partial^\m W^+_\m +{i\over 2} g_2 v \xi_+ \sum_{i} C_{W^+,i} {\cal G}_i \Big)
\Big(\partial^\m W^-_\m - {i\over 2}g_2 v \xi_- \sum_{i} C_{W^-,i} {\cal G}_i \Big).
\nn\eea
Here, we want to raise some points concerning the gauge-fixing procedure:
\bi
\item Thanks to this Lagrangian, all couplings \eqref{terms to be removed} generated by the Goldstone
modes are removed from the action.
\item The gauge fields carry an extra term $-{1\over 2 \xi} (\partial_\m X^\m)^2$ which is inserted in their propagator so that
\bea
&&D_{X}^{\m\n} (k)= {i\over k^2-m_{X}^2} \Big( g^{\m\n} - (1-\x){k^\m k^\n\over k^2-\x m_{X}^2} \Big)
\eea
where $D_{X}/m_{X}$ the propagator and the mass of the gauge field $X^\m$.

\item The ghosts associated with $U(1)_Y\times U(1)_A$ decouple from the rest (there is no coupling between ghosts and gauge fields); therefore, we can absorb the $U(1)_Y\times U(1)_A$ ghost partition function to an overall factor of the total partition function.

\item The most convenient gauge for our purpose is the unitary gauge ($\x_i\to \infty$), where the anomalous gauge fields absorb the Goldstone modes.
Therefore, in the broken phase (unitary gauge), all Goldstone modes have been absorbed by the gauge fields, and we remain with
\bi
\item Massive fermions (quarks, leptons).
\item A massive scalar (massive Higgs - with $m=125$ GeV.).
\item A massless gauge boson (photon $\g$).
\item Two neutral massive bosons ($Z_0$, $Z'$).
\item Two charged massive bosons ($W^\pm$).
\item Gluons.
\ei
\ei
To obtain the couplings of matter fields and gauge fields in the broken phase, we can use the unbroken couplings and make the change
\bea
G_\m^a,~W_\m^a,~Y_\m,~A_\m ~~~~\to ~~~~~  G_\m^a,~W_\m^\pm,~Z_\m,~Z'_\m,~\gamma_\m
\eea
Therefore, from the GCS terms in \eqref{Lextrafields}, can appear ${3\times 4\times 5 \over 1\times 2 \times 3}=10 $ different terms. However, many cancel due to the antisymmetry of the GCS. Finally, we obtain the following GCS terms
\bea
&&Z \g \g ~,~~Z' \g \g~,~~
\g Z Z  ~,~~Z' Z Z~,~~
\g Z'Z'~,~~ Z Z' Z' ~,~~\g Z Z'~~~~
\label{Eq:vertex}
\eea
{\it None of these terms} are present in the classical computation
of the $g-2$, and we shall show that they can be important in some regions of the parameter space. More importantly, their strengths
are determined by the anomaly cancellation conditions,
unveiling informations on the UV particle contents.

\section{The triangle diagram}\label{The tiangle diagram}

In this section, we shall discuss the properties of the effective vertex coming from the triangle diagram.
As it is shown in the appendix, we can split the triangle diagram of a single fermion in the loop as
\bea
\scalebox{0.7}{
\begin{tikzpicture}[thick,baseline={-0.1cm}]
  \begin{feynman}[every blob={/tikz/fill=gray!30,/tikz/inner sep=2pt}]
    \vertex (f1) at (0.75,1){\(i\)};
    \vertex (f2) at (0.75,-1){\(j\)};
    \vertex (ff1) at (1.6125,0.5);
    \vertex (ff2) at (1.6125,-0.5);
    \vertex (f3) at (2.5,0);
    \vertex (e1) at (3.5,0){\(k\)};
    \diagram* {
      (ff1) -- [boson] (f1),
      (ff2) -- [boson] 
      (f2),
      (f3) -- [fermion, edge label'=\(\textrm{$\psi_f$}\)] (ff1),
      (ff2) -- [fermion] (f3),
      (ff1) -- [fermion] (ff2),
      (e1) -- [boson] (f3) }; \end{feynman}
\end{tikzpicture}}
&=&
\Big\{ q^{f-A_i} q^{f-A_j} q^{f-A_k} \Big\}
~~ {\cal C }^{(full)}_\triangle[p_i,p_j,p_k]
\nn\\
&&\raisebox{-0.9mm}[0pt][0pt]{$
=\Big\{ q^{f-A_i} q^{f-A_j} q^{f-A_k} \Big\}
\bigg({\cal C }^{(0)}_\textrm{$\triangle$}[p_i,p_j,p_k] + {\cal C }_\textrm{$\triangle$}^{(m_f)}[p_i,p_j,p_k]\bigg)$}
\nn\eea
where the $q_{V/A}^{f-A_{i/j/k}}$ are the vectrorial/axial charges of the fermion under the gauge bosons $A_{i,j,k}$ respectively,
the $\{q^{f-A_i} q^{f-A_j} q^{f-A_k}\}$ is defined as the collective contribution of the different charges to the triangle diagram
\bea
\Big\{ q^{f-A_i} q^{f-A_j} q^{f-A_k} \Big\}
&=&
q_A^{f-A_i} q_V^{f-A_j} q_V^{f-A_k}+q_V^{f-A_i} q_A^{f-A_j} q_V^{f-A_k}\nn\\
&&
+q_V^{f-A_i} q_V^{f-A_j} q_A^{f-A_k}
+q_A^{f-A_i} q_A^{f-A_j} q_A^{f-A_k}
\eea
and $p_{i,j,k}$ the external momenta and
\bi
\item the $m_f$-independent part of the effective vertex, after taking the limit $m_f\to \infty$, i.e., the part of decoupled fermions:
\bea
{\cal C }^{(0)}_\textrm{$\triangle$}[p_i,p_j,p_k]=\lim_{m_f\to\infty}
{\cal C }^{(full)}_\textrm{$\triangle$}[p_i,p_j,p_k],
\eea
\item the $m_f$-dependent part, which is taken after subtracting the $m_f$ independent part:
\bea
{\cal C }^{(m_f)}_\textrm{$\triangle$}[p_i,p_j,p_k]={\cal C }_\textrm{$\triangle$}^{(full)}[p_i,p_j,p_k]-{\cal C }_\textrm{$\triangle$}^{(0)}[p_i,p_j,p_k].
\eea
\ei
Considering now several fermions $f$ in the triangle loop, the ${\cal C }^{(0)}_\textrm{$\triangle$}$ is common for all of them, and it factorizes out giving the vertex %
\bea
&&\sum_f
\Big\{ q^{f-A_i} q^{f-A_j} q^{f-A_k} \Big\}
\times
 {\cal C }^{(0)}_\textrm{$\triangle$}[p_i,p_j,p_k]
 = t_{ijk} ~ {\cal C }^{(0)}_\textrm{$\triangle$}[p_i,p_j,p_k] \label{theanomalydefinition0}
\eea
We denote by $t_{ijk}$ the anomaly that is related to those defined in equation (\ref{Aanomalies}) on a different basis.
 Notice that since there is an odd number of $\g_5$'s in the anomalous coupling, the trace runs over all possible combinations of charges with one axial and two vectorial.
Therefore, the anomaly is defined as
\bea
t_{ijk}=\sum_f \Big\{
q^{f-A_i}q^{f-A_j} q^{f-A_k}
\Big\}
&=&\sum_f \bigg(
q_A^{f-A_i}q_V^{f-A_j} q_V^{f-A_k}+
q_V^{f-A_i}q_A^{f-A_j} q_V^{f-A_k} \nn\\
&&~~~~~~+ q_V^{f-A_i}q_V^{f-A_j} q_A^{f-A_k}+ q_A^{f-A_i}q_A^{f-A_j} q_A^{f-A_k}\bigg)~~~~
\label{theanomalydefinition}\eea
Thus, the total contribution of the triangle anomaly vertex can be written
\bea
\scalebox{0.7}{
\begin{tikzpicture}[thick,baseline={-0.1cm}]
  \begin{feynman}[every blob={/tikz/fill=gray!30,/tikz/inner sep=2pt}]
    \vertex (f1) at (0.75,1){\(i\)};
    \vertex (f2) at (0.75,-1){\(j\)};
    \vertex (ff1) at (1.6125,0.5);
    \vertex (ff2) at (1.6125,-0.5);
    \vertex (f3) at (2.5,0);
    \vertex (e1) at (3.5,0){\(k\)};
    \diagram* {
      (ff1) -- [boson] (f1),
      (ff2) -- [boson] 
      (f2),
      (f3) -- [fermion, edge label'=\(\textrm{all~$\psi_f$}\)] (ff1),
      (ff2) -- [fermion] (f3),
      (ff1) -- [fermion] (ff2),
      (e1) -- [boson] (f3) }; \end{feynman}
\end{tikzpicture}}=
t_{ijk} {\cal C }^{(0)}_\textrm{$\triangle$}[p_i,p_j,p_k] + \sum_f \Big\{ q^{f-A_i} q^{f-A_j} q^{f-A_k} \Big\} ~{\cal C }_\textrm{$\triangle$}^{(m_f)}[p_i,p_j,p_k]~~~~~
\label{splittingMASSindeANDdePARTS}\eea
where the ellipsis denotes that the axial charge should be taken for all $i,j,k$ and the ${\cal C }^{(m_f)}_\textrm{$\triangle$}$ is different for each massive fermion.
Note that when several fermions are charged under the gauge fields $A_{i,j,k}$, the effective vertex is the sum of all of them.

It is crucial to mention that
\bi
\item In anomaly-free theories, the anomaly is zero $t_{ijk}=0$, and the mass independent part does not contribute.

\item In superficially anomalous theories
like the ones we consider in this work, where the anomaly is canceled by the Green-Schwarz mechanism and GCS terms, the anomalous triangle part contributes accompanied by the axionic and GCS terms necessary to cancel the anomalies.
The effective three-gauge-field coupling will be denoted with a yellow blob with $\cal A$ inside
\bea
&&
\scalebox{0.7}{
\begin{tikzpicture}[thick,baseline={-0.1cm}]
  \begin{feynman}[every blob={/tikz/fill=gray!30,/tikz/inner sep=2pt}]
    \vertex (f1) at (0.75,1);
    \vertex (f2) at (0.75,-1);
    \vertex (e1) at (3.5,0);
    \vertex (g1) [blob,style=yellow] at (1.95,0) {~~~\({\black{\cal A}}\)~~~~};
    \diagram* {
      (f1) -- [boson] (g1),
      (f2) -- [boson]
      (g1),
      (e1) -- [boson] (g1) }; \end{feynman}
\end{tikzpicture}}
=
t_{ijk}\Bigg(
\scalebox{0.7}{
\begin{tikzpicture}[thick,baseline={-0.1cm}]
  \begin{feynman}[every blob={/tikz/fill=gray!30,/tikz/inner sep=2pt}]
    \vertex (f1) at (0.75,1);
    \vertex (f2) at (0.75,-1);
    \vertex (ff1) at (1.6125,0.5);
    \vertex (ff2) at (1.6125,-0.5);
    \vertex (f3) at (2.5,0);
    \vertex (e1) at (3.5,0);
    \vertex (g1)[blob,style=yellow] at (1.95,0) {~~~~~~~~~~~};
    \diagram* {
      (ff1) -- [boson] (f1),
      (ff2) -- [boson] 
      (f2),
      (f3) -- [fermion, edge label'=\(\textrm{$\psi_f$}\)] (ff1),
      (ff2) -- [fermion, edge label'=\(\textrm{$m_f$-independent}\)] (f3),
      (ff1) -- [fermion] (ff2),
      (e1) -- [boson] (f3) }; \end{feynman}
\end{tikzpicture}}
~+~
\scalebox{0.7}{
\begin{tikzpicture}[thick,baseline={-0.1cm}]
  \begin{feynman}[every blob={/tikz/fill=gray!30,/tikz/inner sep=2pt}]
    \vertex (f1) at (0.975,0.75);
    \vertex (f2) at (0.975,-0.75);
    \vertex (ff1) at (1.6125,0.375);
    \vertex (ff2) at (1.6125,-0.320);
    \vertex (f3) at (1.95,0);
    \vertex (e1) at (3.25,0);
    \vertex (g1)[blob,style=yellow] at (1.9,0) {~~~~~~~~~~~};
    \diagram* {
      (f3) -- [boson] (f1),
      (ff2) -- [boson]
      (f2),
      (ff2) -- [scalar] (f3),
      (e1) -- [boson] (f3) }; \end{feynman}
\end{tikzpicture}
}
~+~
\scalebox{0.7}{
\begin{tikzpicture}[thick,baseline={-0.1cm}]
  \begin{feynman}[every blob={/tikz/fill=gray!30,/tikz/inner sep=2pt}]
    \vertex (f1) at (0.975,0.75);
    \vertex (f2) at (0.975,-0.75);
    \vertex (ff1) at (1.6125,0.375);
    \vertex (ff2) at (1.6125,-0.375);
    \vertex (f3) at (1.95,0);
    \vertex (e1) at (3.25,0);
    \vertex (g1)[blob,style=yellow] at (1.9,0) {~~~~~~~~~~~};
    \diagram* {
      (f3) -- [boson
      ] (f1),
      (f3) -- [boson] 
      (f2),
      (e1) -- [boson] (f3) }; \end{feynman}
\end{tikzpicture}}
\Bigg)~~~~
\label{The anomalous effective vertex}\eea
and contains respectively the $m_f$-independent part of the triangle diagram \eqref{splittingMASSindeANDdePARTS} ${\cal C }^{(0)}_\textrm{$\triangle$}[p_i,p_j,p_k]$, the axionic and the GCS vertices.

\item {

The value of $t_{ijk}$ can be huge. Either if all SM fermions contribute additive to the anomaly or if some of these fermions have substantial charges. Both cases are possible as soon as they do not violate any experimental bounds for the couplings $ g_{A} q^{f-A}$.

For example, assuming that all SM fermions have the same charge under the additional $U(1)_A$, we find for some indicative values
\bea
\ba{lll}q^{f-A}_L=2.6\\
q^{f-A}_R=2.0 \ea~ \to~t_{AAA}=100
~~~,~~~~~~~
\ba{lll}q^{f-A}_L=4.5\\
q^{f-A}_R=3.7 \ea~ \to~t_{AAA}=500\nn
\eea
where $q^{f-A}_{L/R}=q^{f-A}_V\pm q^{f-A}_A$ and we keep the relation between vectorial and axial charges to be $q^{f-A}_V=10 q^{f-A}_A$ for purposes explained in the next sections. Relaxing this relation, we can obtain even larger values for the anomaly $t_{AAA}$.}

\ei
The effective vertex \eqref{The anomalous effective vertex}
appears in anomalous theories even when none of the external bosons is the (anomalous) $Z'$. For example, in the model studied in the previous sections, there are such vertices between $\g\g Z$ and $\g ZZ$ as has been shown in section \ref{GCS in mass eigenstate basis}.
Note also that, even if the anomalous vertex \eqref{The anomalous effective vertex} looks like being at tree-level (the last two vertices in the figure above), the axionic and GCS
contributions are of the same order
as the one-loop diagrams (the triangle)
to ensure the anomaly cancellation
condition (\ref{anomalyConditionAY}).
Therefore, any diagram that contains this vertex is of one-loop order more than what is effectively presented.
We now have all the tools to evaluate the contribution of an anomalous $Z'$ and the new effective vertices to the $g-2$ of the muon.

\section{Contributions to $g-2$}\label{g-2}

Now that we have built our Lagrangian in the electroweak broken phase, we can evaluate the contribution of an anomalous gauge
field $Z'_\mu$ to the $g-2$. Several diagrams contribute to the muon anomalous moment, but only a few vertices in (\ref{Eq:vertex}) dominate the process.

\subsection{The diagrams}

Typically, the $g-2$ contribution arises from radiative corrections to the vertex diagram
\bea
\scalebox{0.7}{
\begin{tikzpicture}[thick,baseline={-0.1cm}]
  \begin{feynman}[every blob={/tikz/fill=gray!30,/tikz/inner sep=2pt}]    \vertex[blob] (f0) at (0.5,0) {~~~~~1PI~~~~~};
    \vertex (i1) at (-1.25, 1.5) {\(\bar u\)};
    \vertex (i2) at (-1.25,-1.5){\(u\)};
    \vertex (f1) at (-0.1,0.75);
    \vertex (f2) at (-0.1,-0.75);
    \vertex (f3) at (1.45,0);
    \vertex (e1) at (3,0) {\(\m\)} ;

    \diagram* {
      (f1) -- [fermion, momentum'=\(p'\)] (i1),
      (i2) -- [fermion, momentum'=\(p\)] (f2),
      (e1) -- [boson,
      momentum'=\(q\)] (f3) }; \end{feynman}
\end{tikzpicture}}
&=& (-i e)\bar u(p') \Big( \g^\m F_1(q) +i {\s^{\m\n}q_\n \over 2 m_\ell}F_2(q)
\label{g-2 and EDM}\\
&&~~~~~~~~~~~~~~~~
\raisebox{3mm}[0pt][0pt]{$+
\g^5{\s^{\m\n}q_\n \over 2 m_\ell}F_3(q)
+\g^5(q^2 \g^\m-\sla q q^\m)F_4(q)
\Big) u(p) $}\nn
\eea
where $F_1,F_2,F_3,F_4$ are the contributions to the charge renormalization, $g-2$, EDM and anapole moment respectively.
Especially, $a_\m={1\over 2} (g-2)=F_2(0)$.

Quantum corrections include the propagation of several fields (from the SM sector and the $Z'$), creating various loops.
We can split the extra diagrams that contribute to $g-2$ into three categories:
\bi
\item Diagrams that contain only SM fields and SM vertices.

These diagrams look like
\bea
\scalebox{0.7}{
\begin{tikzpicture}[thick,baseline={-0.1cm}]
  \begin{feynman}[every blob={/tikz/fill=gray!30,/tikz/inner sep=2pt}]
    \vertex (i1) at (-0.75, 1.2);
    \vertex (i2) at (-0.75,-1.2);
    \vertex (f1) at (0,0.75);
    \vertex (f2) at (0,-0.75);
    \vertex (f3) at (1.25,0);
    \vertex (e1) at (2.5,0) {\(\m\)} ;
    \diagram* {
      (f1) -- [fermion] (i1),
      (f2) -- [boson, 
      edge label=\(\textrm{$\g$, $Z$, $W^\pm$}\)] (f1),
      (i2) -- [fermion] (f2),
      (f3) -- [fermion] (f1),
      (f2) -- [fermion] (f3),
      (e1) -- [boson,
      momentum'=\(q\)] (f3) }; \end{feynman}
\end{tikzpicture}}
~+~
\scalebox{0.7}{
\begin{tikzpicture}[thick,baseline={-0.1cm}]
  \begin{feynman}[every blob={/tikz/fill=gray!30,/tikz/inner sep=2pt}]
    \vertex (i1) at (-0.75, 1.2);
    \vertex (i2) at (-0.75,-1.2);
    \vertex (f1) at (0,0.75);
    \vertex (f2) at (0,-0.75);
    \vertex (f3) at (1.25,0);
    \vertex (e1) at (2.5,0) {\(\m\)} ;
    \diagram* {
      (f1) -- [fermion] (i1),
      (f2) -- [scalar, edge label=\(\textrm{Higgs}\)] (f1),
      (i2) -- [fermion] (f2),
      (f3) -- [fermion] (f1),
      (f2) -- [fermion] (f3),
      (e1) -- [boson,
      momentum'=\(q\)] (f3) }; \end{feynman}
\end{tikzpicture}}
~+~
\scalebox{0.7}{
\begin{tikzpicture}[thick,baseline={-0.1cm}]
  \begin{feynman}[every blob={/tikz/fill=gray!30,/tikz/inner sep=2pt}]
    \vertex (i1) at (-0.75, 1.2);
    \vertex (i2) at (-0.75,-1.2);
    \vertex (f1) at (0,0.75);
    \vertex (f2) at (0,-0.75);
    \vertex (f3) at (1.25,0);
    \vertex (e1) at (2.5,0) {\(\m\)} ;
    \diagram* {
      (f1) -- [fermion] (i1),
      (f2) -- [fermion] (f1),
      (i2) -- [fermion] (f2),
      (f3) -- [boson,edge label'=\(\textrm{$W^\pm$}\)] (f1),
      (f2) -- [boson] (f3),
      (e1) -- [boson,
      momentum'=\(q\)] (f3) }; \end{feynman}
\end{tikzpicture}}
~+~\textrm{...}~~~
\eea
Notice that at two loops, we also have the contribution from the fermionic triangle sub-diagram
\bea
\scalebox{0.7}{
\begin{tikzpicture}[thick,baseline={-0.1cm}]
  \begin{feynman}[every blob={/tikz/fill=gray!30,/tikz/inner sep=2pt}]
    \vertex (i1) at (0, 1.5);
    \vertex (i2) at (0,-1.5);
    \vertex (f1) at (0.75,1);
    \vertex (f2) at (0.75,-1);
    \vertex (ff1) at (1.6125,0.5);
    \vertex (ff2) at (1.6125,-0.5);
    \vertex (f3) at (2.5,0);
    \vertex (e1) at (3.5,0) {\(\m\)};
    \vertex (g1) at (1.95,0);
    \diagram* {
      (f1) -- [fermion] (i1),
      (f2) -- [fermion] (f1),
      (i2) -- [fermion] (f2),
      (ff1) -- [boson, edge label'=\(\textrm{$Z$}\)
      ] (f1),
      (ff2) -- [boson, edge label=\(\textrm{$\g$}\)
      ] (f2),
      (f3) -- [fermion, edge label'=\(\textrm{$\psi_f$}\)] (ff1),
      (ff2) -- [fermion] (f3),
      (ff1) -- [fermion] (ff2),
      (e1) -- [boson] (f3) }; \end{feynman}
\end{tikzpicture}}
+
\scalebox{0.7}{
\begin{tikzpicture}[thick,baseline={-0.1cm}]
  \begin{feynman}[every blob={/tikz/fill=gray!30,/tikz/inner sep=2pt}]
    \vertex (i1) at (0, 1.5);
    \vertex (i2) at (0,-1.5);
    \vertex (f1) at (0.75,1);
    \vertex (f2) at (0.75,-1);
    \vertex (ff1) at (1.6125,0.5);
    \vertex (ff2) at (1.6125,-0.5);
    \vertex (f3) at (2.5,0);
    \vertex (e1) at (3.5,0) {\(\m\)};
    \vertex (g1) at (1.95,0);
    \diagram* {
      (f1) -- [fermion] (i1),
      (f2) -- [fermion] (f1),
      (i2) -- [fermion] (f2),
      (ff1) -- [boson, edge label'=\(\textrm{$\g$}\)
      ] (f1),
      (ff2) -- [boson, edge label=\(\textrm{$Z$}\)
      ] (f2),
      (f3) -- [fermion, edge label'=\(\textrm{$\psi_f$}\)] (ff1),
      (ff2) -- [fermion] (f3),
      (ff1) -- [fermion] (ff2),
      (e1) -- [boson] (f3) }; \end{feynman}
\end{tikzpicture}}
+\scalebox{0.7}{
\begin{tikzpicture}[thick,baseline={-0.1cm}]
  \begin{feynman}[every blob={/tikz/fill=gray!30,/tikz/inner sep=2pt}]
    \vertex (i1) at (0, 1.5);
    \vertex (i2) at (0,-1.5);
    \vertex (f1) at (0.75,1);
    \vertex (f2) at (0.75,-1);
    \vertex (ff1) at (1.6125,0.5);
    \vertex (ff2) at (1.6125,-0.5);
    \vertex (f3) at (2.5,0);
    \vertex (e1) at (3.5,0) {\(\m\)};
    \vertex (g1) at (1.95,0);
    \diagram* {
      (f1) -- [fermion] (i1),
      (f2) -- [fermion] (f1),
      (i2) -- [fermion] (f2),
      (ff1) -- [boson, edge label'=\(\textrm{$Z$}\)
      ] (f1),
      (ff2) -- [boson, edge label=\(\textrm{$Z$}\)
      ] (f2),
      (f3) -- [fermion, edge label'=\(\textrm{$\psi_f$}\)] (ff1),
      (ff2) -- [fermion] (f3),
      (ff1) -- [fermion] (ff2),
      (e1) -- [boson] (f3) }; \end{feynman}
\end{tikzpicture}}
\label{triangleSubdiagramSM}\eea
which depend on the mass of the SM fermion in the triangle loop and the individual couplings of the fermions under the $\g,Z$. As we already mentioned, if the theory is anomaly free, the $m_f$ independent part of the triangle diagrams cancels out after summing over all SM fermions in the loop.
The contribution of these diagrams gives the SM prediction for the $g-2$ \eqref{eq:AMMth} \cite{Aoyama:2020ynm}.

{

The couplings between SM matter fields and $\gamma$ and $Z$ are defined in the broken phase of the model. Pure SM and SM with an additional $A$ (or any other extension of the SM) have different expressions for the couplings in the broken phase due to
the contribution of the extra fields.
However, the values of these effective couplings are fixed to the experimental data in terms of the couplings that appear in the unbroken phase. Thus, the values of the minimal couplings between SM fermions and $\gamma$ and $Z$ are independent of
the underline theory difference in the unbroken phase. The values of the tree level couplings of the unbroken theory, do depend on the realization.

}

\item Diagrams that contain the $Z'$ but {\it not} the anomalous coupling.

Such diagrams can be obtained from diagrams of the previous set that contain propagating $Z$ bosons by replacing at least one of them $Z\to Z'$.
At one loop, we have one diagram (where the red vector propagator denotes the $Z'$)
\bea
\scalebox{0.7}{
\begin{tikzpicture}[thick,baseline={-0.1cm}]
  \begin{feynman}[every blob={/tikz/fill=gray!30,/tikz/inner sep=2pt}]
    \vertex (i1) at (-0.75, 1.2);
    \vertex (i2) at (-0.75,-1.2);
    \vertex (f1) at (0,0.75);
    \vertex (f2) at (0,-0.75);
    \vertex (f3) at (1.25,0);
    \vertex (e1) at (2.5,0) {\(\m\)} ;
    \diagram* {
      (f1) -- [fermion] (i1),
      (f2) -- [boson, style=red,
      edge label=\(\textrm{$Z'$}\)] (f1),
      (i2) -- [fermion] (f2),
      (f3) -- [fermion] (f1),
      (f2) -- [fermion] (f3),
      (e1) -- [boson,
      momentum'=\(q\)] (f3) }; \end{feynman}
\end{tikzpicture}}
\label{Fig:treelevel}
\eea
and at two loops, we have
\bea
\scalebox{0.7}{
\begin{tikzpicture}[thick,baseline={-0.1cm}]
  \begin{feynman}[every blob={/tikz/fill=gray!30,/tikz/inner sep=2pt}]
    \vertex (i1) at (-1.25, 1.5);
    \vertex (i2) at (-0.75,-1.2);
    \vertex (g1) at (-0.625,1.125);
    \vertex (g2) at (0.625,0.375);
    \vertex (g3) at (-0.625,0.375);
    \vertex (f1) at (0,0.75);
    \vertex (f2) at (0,-0.75);
    \vertex (f3) at (1.25,0);
    \vertex (e1) at (2.25,0) {\(\m\)} ;
    \diagram* {
      (g1) -- [fermion] (i1),
      (f1) -- [fermion] (g1),
      (f2) -- [boson, 
      edge label=\(\textrm{$\g$, $Z$, $W^\pm$}\)] (f1),
      (i2) -- [fermion] (f2),
      (f3) -- [fermion] (g2),
      (g2) -- [fermion] (f1),
      (f2) -- [fermion] (f3),
      (g1) -- [boson, half left,style=red,edge label=\(\textrm{$Z'$}\)] (g2),
      (e1) -- [boson,
      momentum'=\(q\)] (f3) }; \end{feynman}
\end{tikzpicture}}
~+~
\scalebox{0.7}{
\begin{tikzpicture}[thick,baseline={-0.1cm}]
  \begin{feynman}[every blob={/tikz/fill=gray!30,/tikz/inner sep=2pt}]
    \vertex (i1) at (-0.75, 1.2);
    \vertex (i2) at (-0.75,-1.2);
    \vertex (g1) at (-0.625,1.125);
    \vertex (g2) at (0.625,0.375);
    \vertex (g3) at (0.625,-0.375);
    \vertex (f1) at (0,0.75);
    \vertex (f2) at (0,-0.75);
    \vertex (f3) at (1.25,0);
    \vertex (e1) at (2.25,0) {\(\m\)} ;
    \diagram* {
      (f1) -- [fermion] (i1),
      (f2) -- [boson, 
      edge label=\(\textrm{$\g$, $Z$, $W^\pm$}\)] (f1),
      (i2) -- [fermion] (f2),
      (f3) -- [fermion] (g2),
      (g2) -- [fermion] (f1),
      (f2) -- [fermion] (g3),
      (g3) -- [fermion] (f3),
      (g2) -- [boson,style=red] (g3),
      (e1) -- [boson,
      momentum'=\(q\)] (f3) }; \end{feynman}
\end{tikzpicture}}
~+~\textrm{...}~~~
\label{Z' anomalous-nonanomalous contributions}
\eea
as well as the fermionic triangle sub-diagram (similar to \eqref{triangleSubdiagramSM}) with at least one of the internal bosons to be the $Z'$
\bea
\scalebox{0.7}{
\begin{tikzpicture}[thick,baseline={-0.1cm}]
  \begin{feynman}[every blob={/tikz/fill=gray!30,/tikz/inner sep=2pt}]
    \vertex (i1) at (0, 1.5);
    \vertex (i2) at (0,-1.5);
    \vertex (f1) at (0.75,1);
    \vertex (f2) at (0.75,-1);
    \vertex (ff1) at (1.6125,0.5);
    \vertex (ff2) at (1.6125,-0.5);
    \vertex (f3) at (2.5,0);
    \vertex (e1) at (3.5,0) {\(\m\)};
    \vertex (g1) at (1.95,0);
    \diagram* {
      (f1) -- [fermion] (i1),
      (f2) -- [fermion] (f1),
      (i2) -- [fermion] (f2),
      (ff1) -- [boson, style=red,edge label'=\(\textrm{$Z'$}\)
      ] (f1),
      (ff2) -- [boson, edge label=\(\textrm{$\g/Z$}\)
      ] (f2),
      (f3) -- [fermion, edge label'=\(\textrm{$\psi_f$}\)] (ff1),
      (ff2) -- [fermion] (f3),
      (ff1) -- [fermion] (ff2),
      (e1) -- [boson] (f3) }; \end{feynman}
\end{tikzpicture}}
+
\scalebox{0.7}{
\begin{tikzpicture}[thick,baseline={-0.1cm}]
  \begin{feynman}[every blob={/tikz/fill=gray!30,/tikz/inner sep=2pt}]
    \vertex (i1) at (0, 1.5);
    \vertex (i2) at (0,-1.5);
    \vertex (f1) at (0.75,1);
    \vertex (f2) at (0.75,-1);
    \vertex (ff1) at (1.6125,0.5);
    \vertex (ff2) at (1.6125,-0.5);
    \vertex (f3) at (2.5,0);
    \vertex (e1) at (3.5,0) {\(\m\)};
    \vertex (g1) at (1.95,0);
    \diagram* {
      (f1) -- [fermion] (i1),
      (f2) -- [fermion] (f1),
      (i2) -- [fermion] (f2),
      (ff1) -- [boson, edge label'=\(\textrm{$\g/Z$}\)
      ] (f1),
      (ff2) -- [boson, style=red, edge label=\(\textrm{$Z'$}\)
      ] (f2),
      (f3) -- [fermion, edge label'=\(\textrm{$\psi_f$}\)] (ff1),
      (ff2) -- [fermion] (f3),
      (ff1) -- [fermion] (ff2),
      (e1) -- [boson] (f3) }; \end{feynman}
\end{tikzpicture}}
+\scalebox{0.7}{
\begin{tikzpicture}[thick,baseline={-0.1cm}]
  \begin{feynman}[every blob={/tikz/fill=gray!30,/tikz/inner sep=2pt}]
    \vertex (i1) at (0, 1.5);
    \vertex (i2) at (0,-1.5);
    \vertex (f1) at (0.75,1);
    \vertex (f2) at (0.75,-1);
    \vertex (ff1) at (1.6125,0.5);
    \vertex (ff2) at (1.6125,-0.5);
    \vertex (f3) at (2.5,0);
    \vertex (e1) at (3.5,0) {\(\m\)};
    \vertex (g1) at (1.95,0);
    \diagram* {
      (f1) -- [fermion] (i1),
      (f2) -- [fermion] (f1),
      (i2) -- [fermion] (f2),
      (ff1) -- [boson, style=red, edge label'=\(\textrm{$Z'$}\)
      ] (f1),
      (ff2) -- [boson, style=red, edge label=\(\textrm{$Z'$}\)
      ] (f2),
      (f3) -- [fermion, edge label'=\(\textrm{$\psi_f$}\)] (ff1),
      (ff2) -- [fermion] (f3),
      (ff1) -- [fermion] (ff2),
      (e1) -- [boson] (f3) }; \end{feynman}
\end{tikzpicture}}
\label{triangleSubdiagramZ'}
\eea
Again, these diagrams depend on the mass $m_f$ and the couplings of the SM fermion with the $\g, Z,Z'$.
In this category, we could include the anomalous coupling \eqref{The anomalous effective vertex}; however, we shall separate it from \eqref{triangleSubdiagramZ'} and include it in the next class of diagrams.

The contributions in \eqref{Fig:treelevel}, \eqref{Z' anomalous-nonanomalous contributions}
and \eqref{triangleSubdiagramZ'}
are present independently of the anomalous or non-anomalous nature of the $Z'$. They have been evaluated in various works in the past
\cite{Leveille:1977rc,Pospelov:2008zw,
Davoudiasl:2014kua,1511.07447,1712.09360,1906.11297,CarcamoHernandez:2019ydc,Jho:2019cxq,Hammad:2021mpl,Chowdhury:2021tnm,Yu:2021suw, Athron:2021iuf,Cadeddu:2021dqx,Bodas:2021fsy,Kriewald:2022erk,DAlise:2022ypp,Panda:2022kbn}. On the other hand, we can argue that the contribution of the diagrams  (\ref{Z' anomalous-nonanomalous contributions})  as well as the mass-dependent part of the diagrams  (\ref{triangleSubdiagramZ'}) are much smaller than the mass-independent part of the (\ref{triangleSubdiagramZ'}) diagrams.

{The diagrams  (\ref{Z' anomalous-nonanomalous contributions}) in the parameter space we are interested in can be estimated as follows, using the results of \cite{Kukhto:1992qv}. When $m_{Z'}\ll m_\mu$, the contribution of \eqref{Z' anomalous-nonanomalous contributions} is
\bea
\delta a_\mu\text{\eqref{Z' anomalous-nonanomalous contributions}}\sim \delta a_\mu^{\rm SM,two-loop} (\frac{m_Z}{m_\mu})^2(\frac{g_{Z'} \sin\theta_W}{e})^2\sim \delta a_\mu^{\rm SM,two-loop}(\frac{g_{Z'}}{10^{-3}})^2
\eea
For $\delta a_\mu^{\rm SM,two-loop}\sim 10^{-10}$ \cite{Kukhto:1992qv},
and taking $g_{Z'} \sim 10^{-4}$ (value chosen for the evaluation of the $(g-2)_\m$) later on) we have
\bea
\delta a_\mu\text{\eqref{Z' anomalous-nonanomalous contributions}}\big|_{g_{Z'}=10^{-4}}\sim 10^{-12}
\label{two-loop estimation}\eea}
The second claim, namely that the mass-dependent parts of the diagrams  (\ref{triangleSubdiagramZ'}) are much smaller than the mass-independent part of the (\ref{triangleSubdiagramZ'}) diagrams will be discussed in section \ref{the anomalous contribution} and explicitly shown in the appendix \ref{Comparing contributions}.

\item Diagrams that are present only if the additional $Z'$ is anomalous.

Such diagrams can be obtained from diagrams of the previous sets that contain at least one fermionic triangle sub-diagram, by replacing this subdiagram with the anomalous vertex \eqref{The anomalous effective vertex}. As we have already mentioned, these diagrams are
proportional to the anomaly $t_{ijk}$ and they are absent in anomaly-free models where $t_{ijk}=0$.
We schematically draw a few examples below,
\bea
%
%
\scalebox{0.75}{
\begin{tikzpicture}[thick,baseline={-0.1cm}]
  \begin{feynman}[every blob={/tikz/fill=gray!30,/tikz/inner sep=2pt}]
    \vertex (i1) at (-0.75, 1.2);
    \vertex (i2) at (-0.75,-1.2);
    \vertex (f1) at (0,0.75);
    \vertex (f2) at (0,-0.75);
    \vertex[blob, style=yellow] (f3) at (1.25,0) {~\({\black{\cal A}}\)~~};
    \vertex (e1) at (2.75,0) {} ;
    \diagram* {
      (f1) -- [fermion] (i1),
      (f2) -- [fermion] (f1),
      (i2) -- [fermion] (f2),
      (f3) -- [boson] (f1),
      (f2) -- [boson,style=red] (f3),
      (e1) -- [boson,
      momentum'=\(q\)] (f3) }; \end{feynman}
\end{tikzpicture}}
+
\scalebox{0.7}{
\begin{tikzpicture}[thick,baseline={-0.1cm}]
  \begin{feynman}[every blob={/tikz/fill=gray!30,/tikz/inner sep=2pt}]
    \vertex (i1) at (-1.25, 1.5);
    \vertex (i2) at (-0.75,-1.2);
    \vertex (g1) at (-0.625,1.125);
    \vertex (g2) at (0.625,0.375);
    \vertex (f1) at (0,0.75);
    \vertex (f2) at (0,-0.75);
    \vertex[blob, style=yellow] (kk) at (0,0) {~\({\black{\cal A}}\)~~};
    \vertex (f3) at (1.25,0);
    \vertex (e1) at (2.25,0);
    \diagram* {
      (g1) -- [fermion] (i1),
      (g2) -- [fermion] (g1),
      (f3) -- [fermion] (g2),
      (f2) -- [fermion] (f3),
      (i2) -- [fermion] (f2),
      (kk) -- [boson] (f2),
      (kk) -- [boson, quarter left, style=red] (g1),
      (kk) -- [boson, quarter right] (g2),
      (e1) -- [boson,
      momentum'=\(q\)] (f3) }; \end{feynman}
\end{tikzpicture}}
+
\scalebox{0.7}{
\begin{tikzpicture}[thick,baseline={-0.1cm}]
  \begin{feynman}[every blob={/tikz/fill=gray!30,/tikz/inner sep=2pt}]
    \vertex (i1) at (-1.25, 1.5);
    \vertex (i2) at (-0.75,-1.2);
    \vertex (g1) at (-0.625,1.125);
    \vertex (gg1) at (0,0);
    \vertex (f1) at (0,0.75);
    \vertex (f2) at (0,-0.75);
    \vertex[blob, style=yellow] (f3) at (1.25,0) {~\({\black{\cal A}}\)~~};
    \vertex (e1) at (2.75,0) {} ;
    \diagram* {
      (g1) -- [fermion] (i1),
      (f1) -- [fermion] (g1),
      (gg1) -- [fermion] (f1),
      (f2) -- [fermion] (gg1),
      (i2) -- [fermion] (f2),
      (f3) -- [boson] (f1),
      (g1) -- [boson, quarter right, style=red] (gg1),
      (f2) -- [boson,style=red] (f3),
      (e1) -- [boson,
      momentum'=\(q\)] (f3) }; \end{feynman}
\end{tikzpicture}}
+\textrm{...}~~~
\label{extra diagrams for g-2}
\eea
Diagrams in this set may or may not contain the anomalous $Z'$. Therefore, we can replace the red propagator (propagating $Z'$) in \eqref{extra diagrams for g-2} with a black one (propagating $\g$ or $Z$), and the diagram will still belong in this category.


The first diagram represents several different diagrams where the internal (red) bosonic propagators can be $\g$, $Z$ or $Z'$. They are eight in total (the diagram with only photons is zero), and they are of the lowest order that contain the effective anomalous vertex ${\cal A}$ \eqref{The anomalous effective vertex}. These diagrams effectively appear at one-loop,  however they are two-loop diagrams. We  focus on them.
Notice also that among these eight diagrams, three have $\g$ and $Z$ in the internal boson lines and consist of only SM fields.

\ei

We now have all the tools to evaluate the contribution to
the $g-2$ of the muon from anomalous $Z'$ gauge bosons,
focusing on the first diagram in
\eqref{extra diagrams for g-2} which is the lowest-order diagram (effectively one-loop) that contains the anomalous coupling\footnote{
It is worth mentioning that the anomalous couplings do not contribute to EDM of the muon since it is CP-even.}.
But before, one should compute the leading contribution,
depicted in Fig.(\ref{Fig:treelevel}).

\subsection{The one-loop contribution}
The one-loop contribution of the $Z'$ boson to the $g-2$
can be written as \cite{Leveille:1977rc,Bodas:2021fsy}
\bea
\d a_{\m}&=& {1\over 4 \p^2}{m^2_{\m}\over m_{Z'}^2} %
g^2_{Z'}
\Bigg(
(q_V^{\m-Z'})^2
{\cal F}_V \Big[{m^2_{\m}\over m_{Z'}^2}\Big]
-
(q_A^{\m-Z'})^2
{\cal F}_A \Big[{m^2_{\m}\over m_{Z'}^2}\Big]
\Bigg)~~~~~
\label{g-2FROMGENERIC}
\eea
where
\bea
&&
{\cal F}_V \Big[{m^2_{\m}\over m_{Z'}^2}\Big]
= \int_0^1 dx {x^2(1-x)\over 1-x+x^2 m^2_{\m}/m^2_{Z'}}
\label{g-2FROMGENERICvectorial}
\\
&&
{\cal F}_A \Big[{m^2_{\m}\over m_{Z'}^2}\Big]
= \int_0^1 dx {x(1-x) (4-x)+2x^3m^2_{\m}/m^2_{Z'}\over 1-x+x^2 m^2_{\m}/m^2_{Z'}}.~~~~~
\label{g-2FROMGENERICaxial}
\eea
We observe that $\delta a_\mu$ is a function of the squares of the vectorial and axial couplings of the muon to the $Z'$,
whereas $m_\m$, $q_{V/A}^{\m-Z'}$
Note that the vectorial and the axial contributions have opposite signs.

{The result \eqref{g-2FROMGENERIC} has been obtained from \cite{Leveille:1977rc,Bodas:2021fsy}. In this work, the authors explicitly mention an ambiguity in such 1-loop diagrams (discussed by Jackiw and Weinberg \cite{Jackiw:1972jz}). However, they also mention that this ambiguity has been resolved in a more rigorous treatment using $R_\xi$ gauge. At the same time, a consistent result has also been obtained when the unitary gauge is used jointly with dimensional regularization
\cite{Bardeen:1972vi}. That procedure was used in the rest of our paper below.}

\subsection{The anomalous coupling contribution}\label{g-2 analysis and results}


The relevant anomalous diagram contributing to the $g-2$ of the muon is
\bea
{\cal A}[A,B] ~ = ~\scalebox{.8}{
\begin{tikzpicture}[thick,baseline={-0.1cm}]
  \begin{feynman}[every blob={/tikz/fill=gray!30,/tikz/inner sep=2pt}]
    \vertex (i1) at (-1.25, 1.5);
    \vertex (i2) at (-1.25,-1.5);
    \vertex (f1) at (0,0.75);
    \vertex (f2) at (0,-0.75);
    \vertex[blob,style=yellow] (f3) at (1.25,0) {\({\black{\cal A}}\)};
    \vertex (e1) at (3.25,0) {\(\m\)} ;
    \diagram* {
      (f1) -- [fermion, momentum'=\(p'\)] (i1),
      (f2) -- [fermion, momentum=\(p-k\)] (f1),
      (i2) -- [fermion, momentum'=\(p\)] (f2),
      (f3) -- [boson, momentum'=\(A(k+q)\)] (f1),
      (f2) -- [boson, momentum'=\(B(k)\)] (f3),
      (e1) -- [boson,
      momentum'=\(\g(q)\)] (f3) }; \end{feynman}
\end{tikzpicture}} +\{A\leftrightarrow B\}
\label{diagram for g-2 and EDM}
\eea
where $A,B$ are any of the $\g, Z, Z'$ bosons.
Moreover, we exclude the diagram with both bosons in the loop to be the photon $A=B=\g$. By the $\{A\leftrightarrow B \}$ we denote the diagram with exchanged $A$ and $B$ bosons when $A$ and $B$ are different. Therefore, the corresponding eight diagrams can be grouped as
\bi
\item Diagrams with at least one $Z'$ propagating in the loop
\bea
{\cal A}[Z',\g]&=&
g_{Z'}g_{EM}\int {d^4k\over (2\p)^4}
\Bigg\{ \bar u(p') (-i \g^a q^{\m-Z'} )
S(k-p)
(-i \g^b q^{\m-\g} )
u(p) \label{diagram for g-2 Z' A}\\
&&
~~~~~~~~~~~~~~~~~~~~~~~~~~ \times D^{Z'}_{a \n} (k+q)
\G_{AYY}^{\n \m \r}[-k-q,q,k] ~
D^{\g}_{\r b}(k)\nn\\
&&
~~~~~~~~~~~~~~~~~~~~
+ \bar u(p') (-i \g^a q^{\m-\g} )
S(k-p)
(-i \g^b q^{\m-Z'} )
u(p) \nn \\
&&
~~~~~~~~~~~~~~~~~~~~~~~~~~ \times D^{\g}_{a \r} (k+q)
~\G_{AYY}^{\n \m \r}[k,q,-k-q] ~
D^{Z'}_{\n b}(k)\Bigg\}~~~~~~~~~\nn\\
%
%
{\cal A}[Z',Z]&=&
g_{Z'}g_{Z}\int {d^4k\over (2\p)^4}
\Bigg\{ \bar u(p') (-i \g^a q^{\m-Z'} )
S(k-p)
(-i \g^b q^{\m-Z} )
u(p) \label{diagram for g-2 Z' Z}\\
&&
~~~~~~~~~~~~~~~~~~~~~~~~~~ \times D^{Z'}_{a \n} (k+q)
\G_{YAA}^{\n \m \r}[-k-q,q,k] ~
D^{Z}_{\r b}(k)\nn\\
&&
~~~~~~~~~~~~~~~~~~
+ \bar u(p') (-i \g^a q^{\m-Z} )
S(k-p)
(-i \g^b q^{\m-Z'} )
u(p) \nn \\
&&
~~~~~~~~~~~~~~~~~~~~~~~~~~ \times D^{Z}_{a \r} (k+q)
~\G_{YAA}^{\n \m \r}[k,q,-k-q] ~
D^{Z'}_{\n b}(k)\Bigg\}~~~~~~~~~\nn\\
%
{\cal A}[Z',Z']&=&
g_{Z'}^2
\int {d^4k\over (2\p)^4}
\Big\{ \bar u(p') (-i \g^a q^{\m-Z'} )
S(k-p)
(-i \g^b q^{\m-Z'} )
u(p)\Big\} \label{diagram for g-2 Z' Z'}\\
&&
~~~~~~~~~~~~~~~~~~~~ \times D^{Z'}_{a \n} (k+q)
~ \G_{YAA}^{\m \n \r}[q,-k-q,k] ~
D^{Z'}_{\r b}(k)~~~~~~~~~\nn
\eea
\item Diagrams without $Z'$ propagating in the loop
\bea
{\cal A}[Z,\g]&=&
g_{Z}g_{EM}
\int {d^4k\over (2\p)^4}
\Bigg\{ \bar u(p') (-i \g^a q^{\m-Z} )
S(k-p)
(-i \g^b q^{\m-\g} )
u(p) \label{diagram for g-2 Z A}\\
&&
~~~~~~~~~~~~~~~~~~~~~~~~~~ \times D^{Z}_{a \n} (k+q)
\G_{AYY}^{\n \m \r}[-k-q,q,k] ~
D^{\g}_{\r b}(k)\nn\\
&&
~~~~~~~~~~~~~~~~~~~
+ \bar u(p') (-i \g^a q^{\m-\g} )
S(k-p)
(-i \g^b q^{\m-Z} )
u(p) \nn \\
&&
~~~~~~~~~~~~~~~~~~~~~~~~~~ \times D^{\g}_{a \r} (k+q)
~\G_{AYY}^{\n \m \r}[k,q,-k-q] ~
D^{Z}_{\n b}(k)\Bigg\}~~~~~~~~~\nn\\
%
%
{\cal A}[Z,Z]&=&
g_{Z}^2
\int {d^4k\over (2\p)^4}
\Big\{ \bar u(p') (-i \g^a q^{\m-Z} )
S(k-p)
(-i \g^b q^{\m-Z} )
u(p)\Big\} \nn \\
&&
~~~~~~~~~~~~~~~~~~~~ \times D^{Z}_{a \n} (k+q)
~ \G_{YAA}^{\m \n \r}[q,-k-q,k] ~
D^{Z}_{\r b}(k)~~~~~~~~~
\label{diagram for g-2 Z Z}
\eea
\ei
where $D'$s are the gauge propagators defined in (\ref{Eq:Dphoton}, \ref{Eq:DZZ'}) and $S (p)$ the fermion propagator (\ref{Eq:S}).
The couplings $\Gamma_{AYY}$ and $\Gamma_{YAA}$ are given in Eqs.(\ref{one-loop-CPodd_tensorAYY_mass}, \ref{axion_tensorAYY}, \ref{GCS_tensorAYY}) and (\ref{one-loop-CPodd_tensorYAA_mass}, \ref{axion_tensorYAA}, \ref{GCS_tensorYAA}) respectively. The charges $q^{f-A}$ contain both the vectorial $q_V^{f-A}$ and axial $q_A^{f-A}$ contributions of any fermions $f$ of the SM (\ref{couplings to leptons}).

Note that in the diagrams (\ref{diagram for g-2 Z' A},\ref{diagram for g-2 Z A}),
we use the coupling $\G_{AYY}$ (\ref{one-loop-CPodd_tensorAYY}, \ref{axion_tensorAYY}, \ref{GCS_tensorAYY}) since two photons are involved in the anomalous coupling. The rest, contain $Z,Z'$ in the loop and the $\G_{YAA}$ (\ref{one-loop-CPodd_tensorYAA}, \ref{axion_tensorYAA}, \ref{GCS_tensorYAA}) is used\footnote{The mass for the $Z$ is coming from an extended Higgs mechanism, where the St\"uckelberg and the Higgs mechanisms contribute the masses of the $Z$ and $Z'$ (see section \ref{Masses of Zs}).}.

After some algebra, it is possible to extract the coefficient $F_2(q)$ of (\ref{g-2 and EDM}) from the amplitudes (\ref{diagram for g-2 Z' A}-\ref{diagram for g-2 Z Z}) whose procedure (performed with Mathematica and the use of package-X \cite{Patel:2015tea}) is tedious and is described in detail in the appendix \ref{Extracting g-2 and EDM}. We have obtained the contribution of the axionic and GCS terms to $g-2$
\bea
\d a_{\m}^\textrm{axion$\&$GCS}&=&
\d a_{\m}^\textrm{axion$\&$GCS}[Z',\g]+
\d a_{\m}^\textrm{axion$\&$GCS}[Z',Z']+
\d a_{\m}^\textrm{axion$\&$GCS}[Z',Z]\nn\\
&&+
\d a_{\m}^\textrm{axion$\&$GCS}[Z,\g]+
\d a_{\m}^\textrm{axion$\&$GCS}[Z,Z]
\label{g-2axionsANDGCS}\eea
where the axionic $\&$ GCS contribution from each diagram (\ref{diagram for g-2 Z' Z}-\ref{diagram for g-2 Z Z}) is listed below
\bea
\d a_{\m}^\textrm{axion$\&$GCS}[Z',\g]&=&
t_{Z',\g,\g}
q_A^{\m-Z'} g_{Z'}^2g_{EM}^2
{m_\m^2 \over 192 \pi^4}
 \int_0^1 dx \int_0^{1-x} dy \label{g-2axionsANDGCS Z' g}\\
&& ~~~~~~~~~~~~~~~~~~~
 \Bigg( {(x + y-2) (2 x + y) \over m_{Z'}^2 (x + y-1)-m_\m^2 y^2 }
 - {(1 + x + y) (2 x + y -2)
 \over  m_{Z'}^2 x + m_\m^2 y^2} \Bigg)\nn\eea
\bea
\d a_{\m}^\textrm{axion$\&$GCS}[Z',Z]&=&-t_{\g,Z',Z} \frac{
q_A^{\m-Z}
q_V^{\m-Z'}}{
q_V^{\m-\g}}
g_{Z}^2g_{Z'}^2
{m_\m^2 \over 192 \pi^4}
 \int_0^1 dx \int_0^{1-x} dy \label{g-2axionsANDGCS Z' Z}\\
&& ~~~~~~~~~~~~~~~~~~~~
\Bigg(
\frac{ (x+y-2)((2 x+y)~{\cal Q}+ 2 x+y-2)}{m_\m^2 y^2+m_{Z}^2 x-m_{Z'}^2 (x+y-1)}\nn\\
&& ~~~~~~~~~~~~~~~~~~~~~~
+\frac{(x+y+1)
((2 x+y-2)~{\cal Q}+ 2 x+y)}{m_\m^2 y^2-m_{Z}^2 (x+y-1)+m_{Z'}^2 x}\nn\\
&&~~~~~~~~
-\frac{1}{3 m_{Z'}^2}\int_0^{1-x-y}dz\bigg( \frac{(m^2_\mu z^2-m_{Z'}^2)}{m_\m^2 z^2-m_{Z}^2 (x+y+z-1)+m_{Z'}^2 y}
\nn\\
&&~~~~~~~~~~~~~~~~~~~~~~~~~~~~~~~~~~~~ -\frac{(m_{Z'}-m_\mu z) (m_\mu z+m_{Z'})}{m_\m^2 z^2+y m_{Z}^2+m_{Z'}^2 (1-x+y+z)}
\bigg)\Bigg)
\nn\eea
\bea
&&\d a_{\m}^\textrm{axion$\&$GCS}[Z',Z']=t_{\g,Z',Z'}
\frac{
q_A^{\m-Z'}
q_V^{\m-Z'}}{3
q_V^{\m-\g}}g_{Z'}^4
{m_\m^2 \over 192 \pi^4}
 \int_0^1 dx \int_0^{1-x} dy \label{g-2axionsANDGCS Z' Z'}\\
&&~~~~~~~~~~~~~~~~~~~~~~
\Bigg(
\frac{(2 x+y-1) (2 x+2 y-1)}{m_\m^2 y^2-m_{Z'}^2 (y-1)}
+\int_0^{1-x-y}dz\frac{3 (m_{Z'}-m_{\m} z) (m_{\m} z+m_{Z'})  }{m_{Z'}^4 (x+z-1)-m_\m^2 m_{Z'}^2 z^2}
\Bigg)\nn\eea
\bea
\d a_{\m}^\textrm{axion$\&$GCS}[Z,\g]&=&
t_{Z,\g,\g}
q_A^{\m-Z} g_{Z}^2 g_{EM}^2
{m_\m^2 \over 192 \pi^4}
 \int_0^1 dx \int_0^{1-x} dy \label{g-2axionsANDGCS Z gamma}\\
&& ~~~~~~~~~~~~~~~~~~~~~
\Bigg( {(x + y-2) (2 x + y) \over m_{Z}^2 (x + y-1)-m_\m^2 y^2 }
 - {(1 + x + y) (2 x + y -2)
 \over  m_{Z}^2 x + m_\m^2 y^2} \Bigg)\nn\eea
\bea
&&\d a_{\m}^\textrm{axion$\&$GCS}[Z,Z]=t_{\g,Z,Z}
\frac{
q_A^{\m-Z}
q_V^{\m-Z}}{3
q_V^{\m-\g}}g_{Z}^4
{m_\m^2 \over 192 \pi^4}
 \int_0^1 dx \int_0^{1-x} dy \label{g-2axionsANDGCS Z Z}\\
&&~~~~~~~~~~~~~~~~~~~~~~~~
\Bigg(
\frac{(2 x+y-1) (2 x+2 y-1)}{m_\m^2 y^2-m_{Z}^2 (y-1)}
+\int_0^{1-x-y}dz\frac{3 (m_{Z}-m_{\m} z) (m_{\m} z+m_{Z})  }{m_{Z}^4 (x+z-1)-m_\m^2 m_{Z}^2 z^2}
\Bigg)\nn
\eea
while the $m_f$ independent part of the triangle diagrams gives
\bea
\d a_\m^\textrm{$\triangle$(mass ind)}
&=&
\d a_\m^\textrm{$\triangle$(mass ind)}[Z',\g]+
\d a_\m^\textrm{$\triangle$(mass ind)}[Z',Z]+
\d a_\m^\textrm{$\triangle$(mass ind)}[Z',Z']\nn\\
&&+\d a_\m^\textrm{$\triangle$(mass ind)}[Z,\g]+
\d a_\m^\textrm{$\triangle$(mass ind)}[Z,Z]
\label{g-2triangleMASSINDEPENDENT}\eea
separated into the different diagrams (\ref{diagram for g-2 Z' Z}-\ref{diagram for g-2 Z Z})
\bea
\d a_\m^\textrm{$\triangle$(mass ind)}[Z',\g]&=&
t_{Z',\g,\g}
q_A^{\m-Z'}{g_{Z'}^2g^2_{EM}}
\frac{ m_\m^2}{576 \pi ^4 }
\label{g-2triangleMASSINDEPENDENT Z' gamma}\\
&& \int_0^1dx\int_0^{1-x}dy \bigg(\frac{(x+y+1) (2 x+y-2)}{m_\m^2 y^2+ m_{Z'}^2 x}
 -\frac{(x+y-2) (2 x+y)}{m_{Z'}^2 (x+y-1)-m_\m^2 y^2}\bigg)\nn\eea
\bea
\d a_\m^\textrm{$\triangle$(mass ind)}[Z',Z]&=&
t_{\g,Z'Z}
{q_A^{\m-Z} q_V^{\m-Z'} \over {
q_A^{\m-\g}}}
{g_{Z}^2g_{Z'}^2}
\frac{ m_\m^2}{576 \pi ^4 }
\label{g-2triangleMASSINDEPENDENT Z' Z}\\
&&\int_0^1dx\int_0^{1-x}dy
\bigg(\frac{(x+y-2) ((2 x+y)~{\cal Q} + 2 x+y-2)}{m_\m^2 y^2+ m_{Z}^2 x - m_{Z'}^2 (x+y-1)}\nn\\
&&~~~~~~~~~~~~~~~~~~~~~~ +\frac{(x+y+1) ((2x+y-2){\cal Q}+
2 x+y)}{m_\m^2 y^2-m_{Z}^2 (x+y-1)+ m_{Z'}^2 x}\bigg)\nn\eea
\bea
\d a_\m^\textrm{$\triangle$(mass ind)}[Z',Z']&=&
-t_{\g,Z',Z'} \frac{
q_A^{\m-Z'} q_V^{\m-Z'}}{q_V^{\m-\g}}
 g^4_{Z'}
\frac{ m_\m^2}{288 \pi ^4 } \label{g-2triangleMASSINDEPENDENT Z' Z'}\\
&&\int_0^1dx\int_0^{1-x}dy \frac{(2 x+y-1) (2 x+2 y-1)}{m_\m^2 y^2- m_{Z'}^2 (y-1)}\nn\eea
\bea
\d a_\m^\textrm{$\triangle$(mass ind)}[Z,\g]&=&
t_{Z,\g,\g}
q_A^{\m-Z}  g^2_{Z} g_{EM}^2
\frac{ m_\m^2}{576 \pi ^4 }\label{g-2triangleMASSINDEPENDENT Z gamma}\\
&& \int_0^1dx\int_0^{1-x}dy \bigg(\frac{(x+y+1) (2 x+y-2)}{m_\m^2 y^2+ m_{Z}^2 x}
 -\frac{(x+y-2) (2 x+y)}{m_{Z}^2 (x+y-1)-m_\m^2 y^2}\bigg)\nn\eea
\bea
\d a_\m^\textrm{$\triangle$(mass ind)}[Z,Z]&=&
-t_{\g,Z,Z} \frac{
 q_A^{\m-Z}q_V^{\m-Z}}{
 q_V^{\m-\g}}
  g^4_{Z}
  \frac{ m_\m^2}{288 \pi ^4 }
\label{g-2triangleMASSINDEPENDENT Z Z}\\
&& \int_0^1dx\int_0^{1-x}dy
\frac{(2 x+y-1) (2 x+2 y-1)}{m_\m^2 y^2- m_{Z}^2 (y-1)}\nn
\eea
where ${\cal Q}=\frac{q_A^{\m-Z'} q_V^{\m-Z}}{q_A^{\m-Z} q_V^{\m-Z'}}$,
$m_{\m/f}$ and $q^{\m/f-{\g/Z/Z'}}_{V/A}$ are the masses and the charges (vectorial/axial) of the muon $\m$ and the lightest SM fermion $f$ charged under $\g,Z,Z'$ respectively.

Finally, the $m_f$ dependant part of the triangle diagrams has a very long expression, which is omitted here.
Even though we try to be generic in this stage of our analysis, we shall use some numerical results to compare each term's contribution.
In appendix \ref{Comparing contributions}, we compare the contributions of each part (the axionic and GCS contribution, the mass-independent part contribution and the mass-dependent part) of the diagrams (\ref{diagram for g-2 Z' A} - \ref{diagram for g-2 Z Z}).
We drop the charges and couplings and focus on the integrals in each case. These contributions depend on the mass of the $Z'$. For reasons explained in the following section, we consider a low mass $Z'$ model with $m_{Z'}=5-100$ MeV, the mass of the muon $m_\m$ and the mass of the $Z$, $m_Z$.
In fig.\ref{Fig:figCall}, we plot the absolute values of these contributions, and we conclude that the leading one is coming from the ${\cal A}[Z',\g]$ and especially from the axionic and GCS parts.

\section{A low mass $Z'$ model and the $g-2$ of the muon}
\label{Sec:ALowMassZ'ModelAndResults}

{

In this section, we apply our results to a low mass $Z'$ model. 
We choose some values for the parameters of the model discussed in section \ref{SM+U(1)+axion} (a discussion of a toy-model and the chosen values is provided in the appendix \ref{An UV completion}), and we evaluate the contribution of an anomalous $Z'$ to the $g-2$ of the muon.

}

\subsection{Justifying our choices}

In the generic theory, the quarks are charged under the $Z'$. In this case, the $Z'$ can be abundantly produced in a hadron collider and the constraints on the $Z'$ mass and coupling are rather stringent, its contribution to the $g-2$ is then relatively small.
The situation becomes more interesting if we suppose that the anomalous $Z'$ couples only to the lepton sector,
{\it i.e.} leptophilic $Z'$, \cite{Buras}.
In that case, we may classify $Z'$ models into two categories, depending on whether $Z'$ couples to the electron (a) or not (b).

\begin{enumerate}
    \item[(a)]
     {In generic models of this category, the couplings to leptons can be independent. As there are strong constraints on the coupling to the electron, as mentioned above, in all such models the contributions to $g-2$ are also small. The only exception is to assume that the quantized charge of the electron is much smaller than that of the muon and tau, a highly unusual situation.}

     A simple example of this class of models is a dark $U(1)$ gauge boson with a kinetic mixing.\footnote{See, e.g., Ref. \cite{Essig:2013lka} for a review.} The kinetic mixing parameter $\epsilon$ may be defined such that the coupling of $Z'$ to the electromagnetic current $J_{\rm em}^\mu$ is given by
    \begin{align}
        {\cal L}_{\rm int} &=
        -\epsilon e Z'_\mu J_{\rm em}^\mu.
    \end{align}
    The muon $g-2$ can be explained at one loop in this model with a parameter space $\epsilon\gtrsim 10^{-3}$ and $m_{Z'}\lesssim 1$ GeV, which is however excluded by the search for a single $\gamma$ event in $e^+e^-\to \gamma Z'$ followed by $Z'\to$ invisible (such as neutrinos) at Babar \cite{BaBar:2017tiz}. In particular, this channel constrains the $Z'$ coupling to the electron as this is t-/u-channel process by exchanging electrons and emitting $Z'$.

    \item[(b)] A typical example of this class of models contains a  $Z'$ that couples to $\mu$ (as well as $\tau$), but to no other fermions of the SM.
     The extra $U(1)$ symmetry is the gauged $U(1)_{L_\mu-L_\tau}$ symmetry. The experimental constraints are discussed in  \cite{Jho:2019cxq} (see also \cite{Kamada:2018zxi}), where the parameter space with $5~{\rm MeV}\lesssim m_{Z'}\lesssim 200~{\rm MeV}$ and the gauge coupling $\sim 5\times10^{-4}$ are compatible with this $Z'$ giving a correction to the muon $g-2$ that has the right size to explain the discrepancy, as in (\ref{Eq:exp}).
\end{enumerate}

It is important to note that in the anomalous $U(1)_A$ models, different flavors may, in principle, have different $U(1)_A$ charges as the anomaly cancellation required in most phenomenological models is not necessary.
Therefore, for instance, the charges under the muon number $U(1)_{L_\mu}$ and the tau number $U(1)_{L_\tau}$ can be different, as opposed to the anomaly-free $U(1)_{L_{\mu}-L_{\tau}}$ model.
We shall come back to this point later.

\subsection{The one-loop contribution}

We have evaluated the contribution of a leptonic $Z'$ that couples only to the muon $\m$ and tau $\t$ fermions of the SM (class (b)). In this case, the lightest fermion that couples to $Z'$ is the muon setting $f=\m$.

We  used the values
\bea
&& m_{\m}=105 ~\textrm{MeV}, ~m_Z=91~\textrm{GeV},\\
&& g_{EM} q_V^{f-\g}=-e=-\sqrt{4\pi\alpha} = -0.30\\
&& g_Z q_V^{f-Z}={e-4 e \sin^2 \theta_W \over 4\sin \theta_W \cos \theta_W}=0.014,\\
&&g_Z q_A^{f-Z}=-{e\over 4\sin \theta_W \cos \theta_W}=-0.18,
\label{some values}\eea
where $\theta_W$ the Weinberg angle with $\sin^2\theta_W = 0.23$, and $\a=1/137$ is the fine structure constant.

\begin{figure}[t]
\begin{center}
\includegraphics[width=120mm]{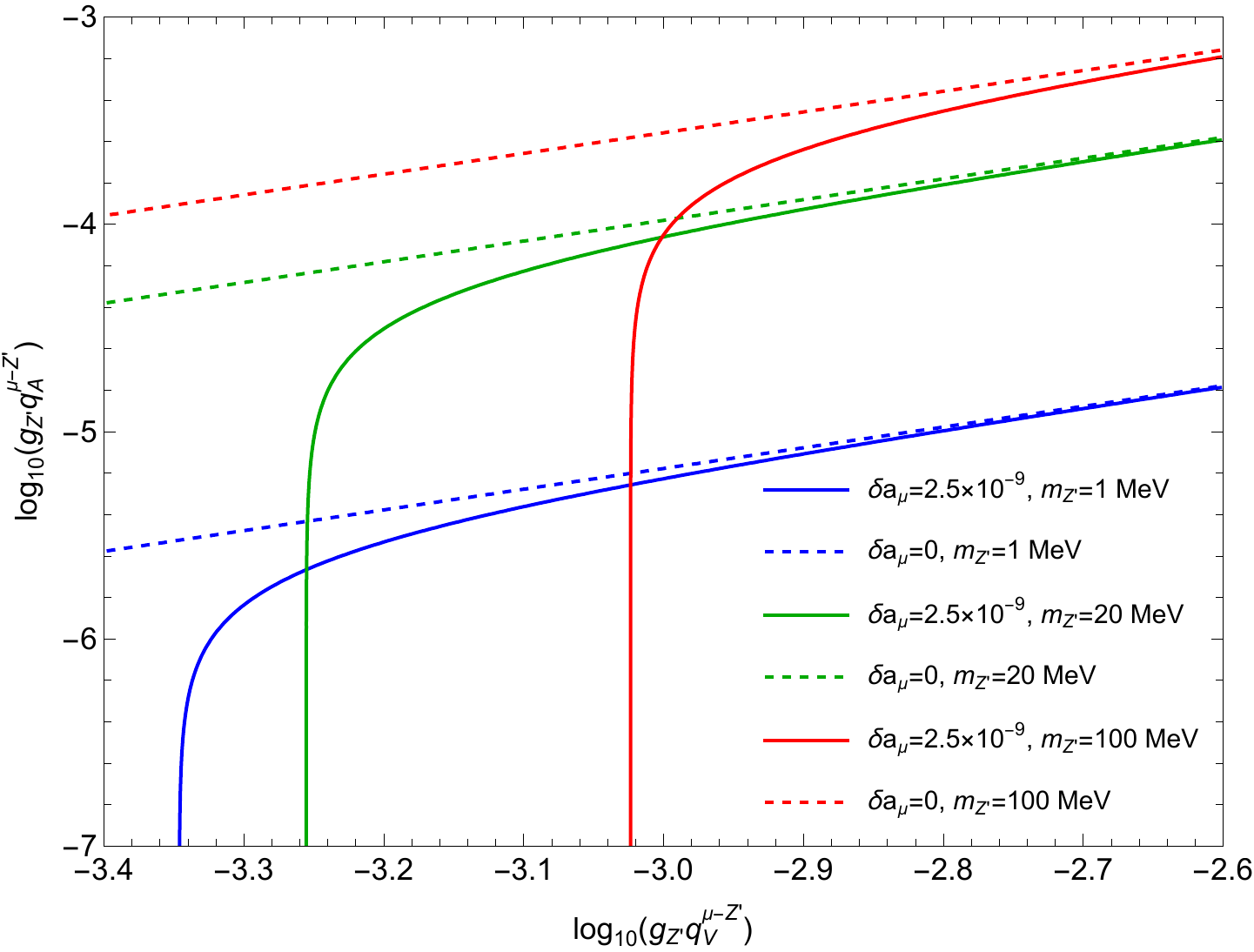}
\caption{
$g_{Z'}q_A^{\m-Z'}$ as a function of $g_{Z'}q_V^{\m-Z'}$, for fixed values of $\d a_\m$. Solid lines have fixed $\d a_\mu=2.5\times 10^{-9}$ \eqref{Eq:exp} and dashed lines have $\d a_\mu=0$ for $m_{Z'}=1, 20,100$ MeV.
}
\label{Fig:qvqa}
\end{center}
\end{figure}

Assuming that the one-loop correction to the muon $g-2$
corresponds to the right magnitude (\ref{Eq:exp}),
from \eqref{g-2FROMGENERIC} we may
obtain the relation between $q_V^{\mu-Z'}$, $q_A^{\mu-Z'}$ and
$m_{Z'}$:
\bea
(g_{Z'} q_A^{\mu-Z'})^2\simeq
{1\over {\cal F}_A[m^2_{\m}/m^2_{Z'}]}
\left(
(g_{Z'}q_V^{\mu-Z'})^2 {\cal F}_V[m^2_{\m}/m^2_{Z'}]- 4\pi^2\frac{m_{Z'}^2}{m_\mu^2} \delta a_\mu\right),
\label{Eq:qvqabis}
\eea
which means that for a given vectorial coupling, one can deduce the axial coupling necessary to fit the experimental constraint.
We show in Fig.(\ref{Fig:qvqa})
the points in the plane ($q_V^{\mu-Z'}$, $q_A^{\mu-Z'}$) respecting the observed
value
$\delta a_\mu=2.5 \times 10^{-9}$
for different values of $m_{Z'}$ (1, 20 and 100 MeV).

For comparison, we also draw the corresponding line, which
ensures a cancellation between the axial and vectorial charges ($\delta a_\mu=0$). The influence
of the axial charge becomes stronger for small
values of $q_V^{\m-Z'}$.
For instance, for $m_{Z'}=20$ MeV, we clearly see an asymptote around
$g_{Z'}q_V^{\m-Z'}=5.6\times 10^{-4}$ for $q_A^{\m-Z'}=0$,
which value is in accordance with the result of \cite{Jho:2019cxq}.
On the other hand,
as $q_V^{\m-Z'}$ increases, the value of $q_A^{\m-Z'}$ needed
to respect, the $g-2$ measurement also increases due
to its negative contribution, reaching $g_{Z'}q_A^{\m-Z'}\simeq 5\times 10^{-5}$ for $g_{Z'}q_V^{\m-Z'}\simeq 10^{-3}$.

To understand the dependence of the $g-2$
on the vectorial and axial charges for
light $Z'$, it is easy to see that
for $m_{Z'} \ll m_\mu$,
Eq.(\ref{Eq:qvqabis}) reduces to
\begin{eqnarray}
(g_{Z'}q_A^{\mu-Z'})^2\simeq \left(\frac{(g_{Z'}q_V^{\mu-Z'})^2}{8\pi^2} - \delta a_\mu\right)\times 4\pi^2\frac{m_{Z'}^2}{m_\mu^2}.
\label{QaFROMQv}
\end{eqnarray}
 This indicates that for $(g_{Z'}q_V^{\mu-Z'})^2\simeq 8\pi^2\delta a_\mu \simeq (5\times10^{-4})^2$ (for the measured value $\delta a_\mu\simeq 3\times10^{-9}$),
 $q_A^{\mu-Z'}\simeq 0$. This is compatible with our
 numerical result obtained in Fig.(\ref{Fig:qvqa}).

 The one-loop contribution can even vanish, i.e. $\delta a_\mu=0$, for $q_V^{\mu-Z'}\simeq (\sqrt{2}m_\mu/m_{Z'})q_A^{\mu-Z'}\simeq 15 \times q^{\mu-Z'}_A$ for $m_{Z'}=10$ MeV. In other words,
for a 10 MeV $Z'$, if
$q_V^{\mu-Z'}\simeq 15\times q_A^{\mu-Z'}$, axial and vectorial contributions cancel,
and higher-order effects, like those generated by the anomalies, should be considered.

\begin{figure}[t]
\begin{center}
\epsfig{file=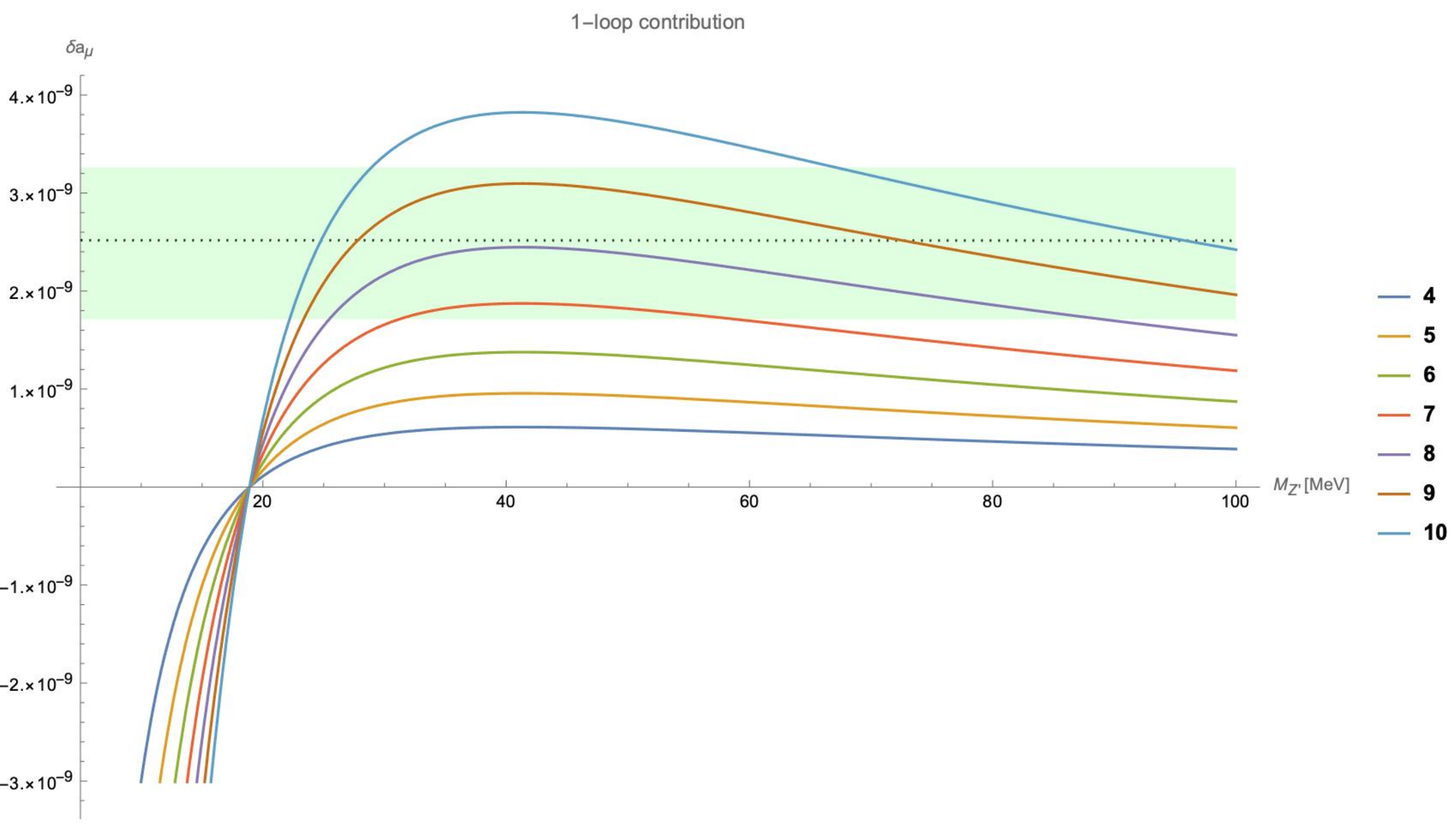,width=140mm}\\
\epsfig{file=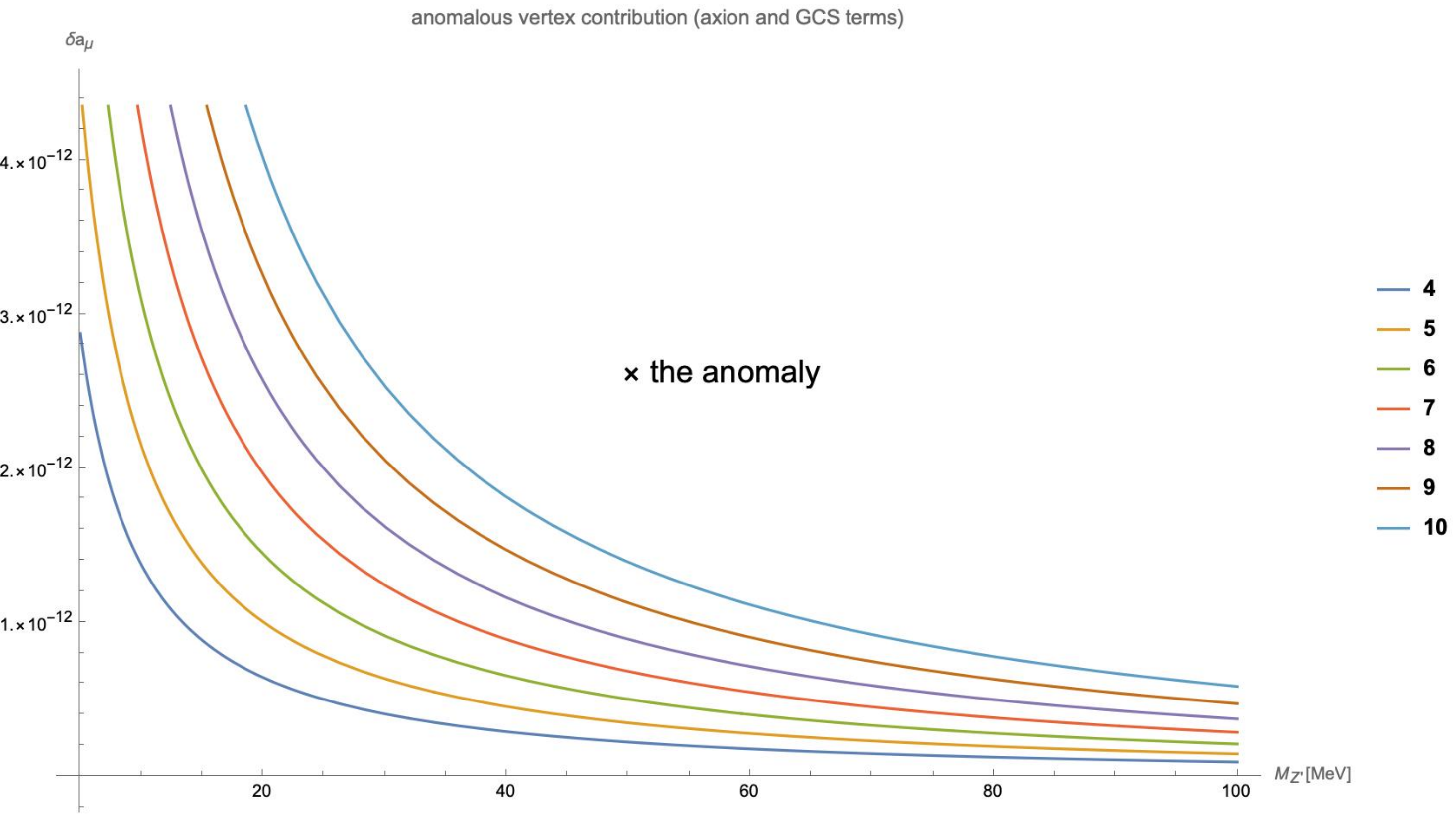,width=140mm}
\caption{The two plots give the contribution to the $g-2$ from the one-loop and the two-loop (anomalous) amplitudes, for different
$g_{Z'}q_V^{\mu-Z'}=10\times g_{Z'} q_A^{\mu-Z'}=(4-10) \times 10^{-4}$,
represented by different colored lines.
In the first plot, the dashed line is the $\delta a$ \eqref{Eq:exp} and the green zone the $1\sigma$ uncertainty ($\d \sigma=\sqrt{63^2+43^2}\times 10^{-11}$).
Notice that the contribution from $Z'$ with masses below the 20 MeV becomes negative.
In the second diagram, we present the contribution of the anomalous diagram {\it without} the overall anomalous factor $t_{ijk}$.
For an anomaly equal to 100, this contribution is within the experimental error and for even larger anomaly values becomes dominant.}
\label{Fig:fig1}
\end{center}
\end{figure}

This behavior
of $g_{Z'}q_A^{\m-Z'}$ is in accordance with Fig.(\ref{Fig:qvqa}),
where we see that the $\delta a_\mu=0$
lines follow $\log(q_A^{\mu-Z'}) = \log(q_V^{\mu-Z'}) + $cst.
The situation is different for heavier $Z'$.
In fact, one can observe from Eq.(\ref{g-2FROMGENERICvectorial}) that if $m_{Z'} \gg m_\mu$, ${\cal F}_V \simeq\frac{1}{3}$, or the axial contribution needed to obtain the right $\delta a_\mu$ vanishes for
\begin{equation}
g_{Z'}q_V^{\mu-Z'}\simeq 6\times 10^{-4}\frac{m_{Z'}}{m_\mu}\sqrt{\frac{\delta a_\mu}{3 \times 10^{-9}}}.
\end{equation}
An interesting feature is also that for a given
pair ($q_V^{\mu-Z'}$,$q_A^{\mu-Z'}$), two values of $m_{Z'}$
are allowed (the lines in Fig.(\ref{Fig:qvqa}) intersect).
This comes from the features of Eq.(\ref{g-2FROMGENERIC}),
where the anomalous moment of the muon first increases with $m_{Z'}$ for $m_{Z'} \ll m_\mu$, whereas it decreases when $m_{Z'} \gg m_\mu$.
We illustrate this behavior on the upper panel of Fig.(\ref{Fig:fig1}) where we
plot $\delta a_\mu$ as function of $m_{Z'}$
for different values of $q_V^{\mu-Z'}$ and
$g_{Z'}q_A^{\mu-Z'}=g_{Z'} q_V^{\mu-Z'}/10$, and the lower panel, the corresponding anomalous contribution of the $Z'$. It is interesting to notice that the contribution of a $Z'$ to $g-2$ reaches a maximum for a $Z'$ mass of $\sim 40$ MeV, much below the electroweak scale.
The $4.2\sigma$ discrepancy (\ref{Eq:exp}) can even be obtained
for $m_{Z'}=40$ MeV and $g_{Z'}(q_V^{\mu-Z'},q_A^{\mu-Z'})$ = ($0.81\times 10^{-3}$, $0.81\times 10^{-4}$).

{For the specific choice of parameters, $g_{Z'}q_V^{\mu-Z'}=10\times g_{Z'} q_A^{\mu-Z'}=(4-10) \times 10^{-4}$, the axial and vectorial contributions arise with an opposite sign, there exists a region of the parameter space $m_{Z'}\simeq 20$ MeV, where both contributions cancel exactly, rendering the contribution generated by the anomalies (\ref{g-2axionsANDGCS}) dominant (fig \ref{Fig:fig1}, upper plot).\footnote{Notice that for different values of $g_{Z'}q_V^{\mu-Z'},g_{Z'} q_A^{\mu-Z'}$, the cancellation in the 1-loop diagrams appears at $m_{Z'}$ different from 20MeV (fig \ref{Fig:qvqa}).}}
The values of the couplings fitting $\delta a_\mu$ are of the order of $\sim 10^{-4}$ and do not enter into conflict with other $Z'$ searches such as the neutrino trident production, which rules out the coupling greater than $10^{-3}$ for $m_{Z'}\lesssim {\cal O}(10)$ MeV \cite{Altmannshofer:2014pba}.

\subsection{The anomaly contribution}\label{the anomalous contribution}


As we have already mentioned, we have three different sets of diagrams depending on the type of the bosons $A$ and $B$ in \eqref{diagram for g-2 and EDM}. These diagrams can be grouped as
\bi
\item[1.] ${\cal A}[Z',\g]$ given in \eqref{diagram for g-2 Z' A}
\item[2.] ${\cal A}[Z',Z]$ given in \eqref{diagram for g-2 Z' Z}
\item[3.] ${\cal A}[Z',Z']$ given in \eqref{diagram for g-2 Z' Z'}
\item[4.] ${\cal A}[Z,\g]$ given in \eqref{diagram for g-2 Z A}
\item[5.] ${\cal A}[Z,Z]$ given in \eqref{diagram for g-2 Z Z}
\ei
Each of these sets of diagrams can be split in
\bi
\item[a.] the axionic $\&$ GCS contribution,
\item[b.] the $m_f$ independent part of the triangle diagram and \item[c.] the $m_f$ dependent part.
\ei
According to \eqref{thecutoff}, the typical cutoff of our model is $\Lambda \sim 40$ TeV, where we took, $m_{Z'}\simeq 20$ MeV, $g_A\simeq 10^{-4}$ and $t_{ijk}\simeq 100$.
This is 1-2 orders of magnitude above the experimental scales. In retrospect, it implies that in such a case there is new physics (new massive charged fermions) within the reach of the next generation of experiments.

Next, we make comparisons between the five different diagrams and the three different contributions per diagram in order to find the leading one.

We start by comparing the contributions from the a, b and c parts in each diagram separately. We drop the overall factor that contains the charges and the couplings, and we focus on the integrals. These formulae depend only on the masses of the fields involved.
We present our results in appendix \ref{Extracting g-2 and EDM}. In all cases, the leading contribution per diagram comes from the axionic and GCS vertices.

Next, comparing the full diagrams (with the couplings and charges) (\ref{diagram for g-2 Z' A} -
\ref{diagram for g-2 Z Z}) we find that the leading contribution to the $g-2$ is coming from ${\cal A}[Z',\g]$ diagram, with one photon and an anomalous $Z'$ in the loop. The rest are suppressed by extra massive gauge fields in the loop (the photon is replaced by $Z$ or $Z'$). Additionally, the contribution of the ${\cal A}[Z,\g]$, ${\cal A}[Z,Z]$ are suppressed by the mass of the $Z$, which is the heaviest boson in our analysis.

\subsubsection*{The leading anomalous contribution}

According to our previous discussion, we focus on the leading contribution coming from the diagram ${\cal A}[Z',\g]$ and especially the axion \& GCS vertex in the loop.
This contribution depends on four parameters. The mass of the $Z'$, $m_{Z'}$, the vectorial/axial couplings of the $Z'$ to the muon $q_{V/A}^{\mu-Z'}$ and the anomaly trace $t_{Z' \g \g}$.
To illustrate our results, we shall scan our parameter
space $m_{Z'}\in (5-100)$ MeV with $g_{Z'}q_V^{\mu-Z'}=10\times g_{Z'} q_A^{\mu-Z'}=(4-10)\times 10^{-4}$.

The overall anomaly $t_{ijk}$ is a free parameter that can take any positive or negative value, depending on the number of fermions charged under the relevant gauge bosons. If we assume that the charges of the fermions under the anomalous $A^\m$ are big (bigger than 4), the anomaly $t_{ijk}$ can be huge.

\begin{figure}[t]
\begin{center}
\epsfig{file=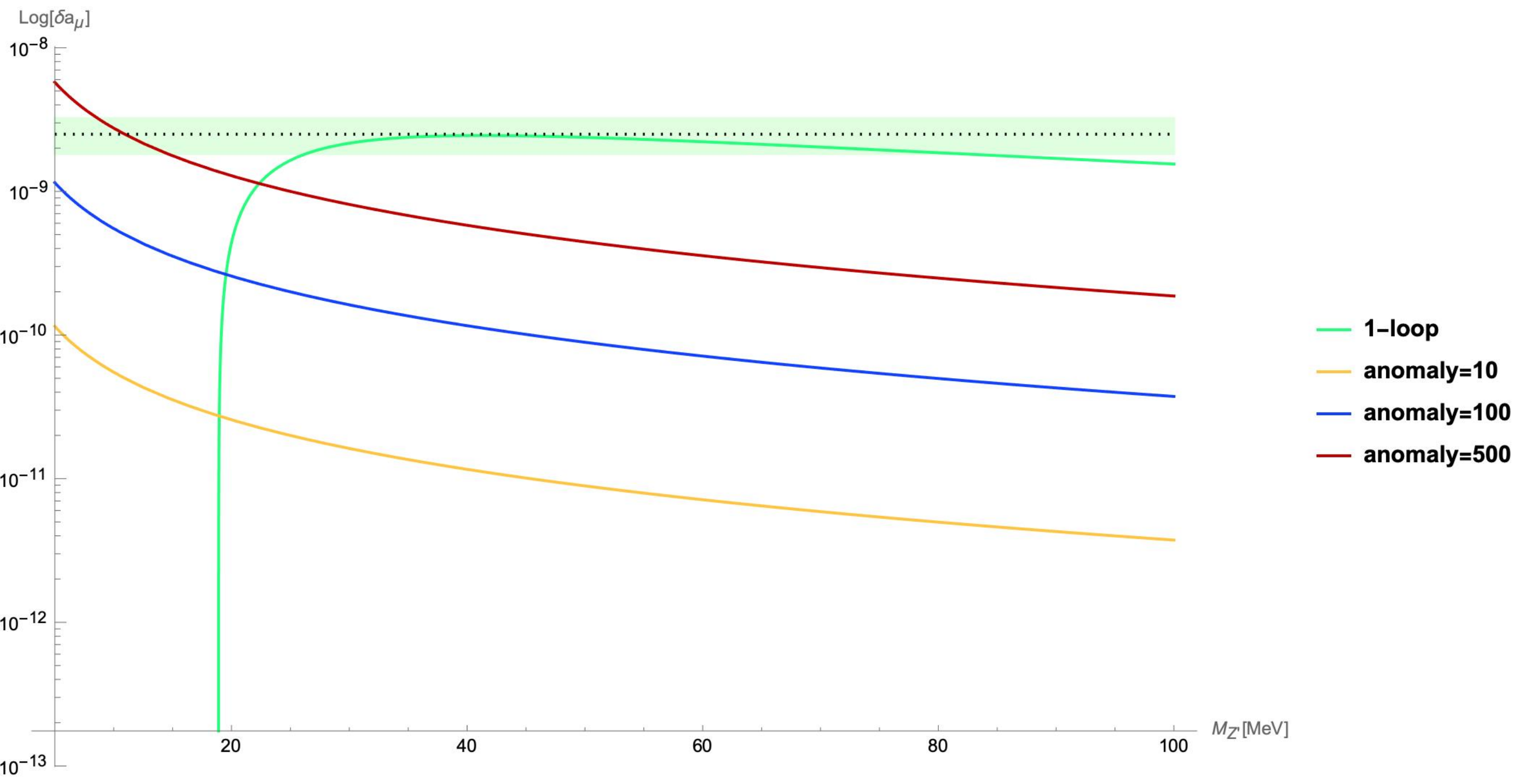,width=145mm}
\caption{In this plot, we present the one-loop contribution of an anomalous $Z'$ to the $g-2$ of the muon for masses $m_{Z'}=5-100$ MeV and fixed couplings
$g_{Z'}q_V^{\mu-Z'}=10\times g_{Z'} q_A^{\mu-Z'}=8\times 10^{-4}$. We also provide the contribution from the anomalous coupling for an anomaly $t_{ijk}=10,100,500$.}\label{Fig:figManyAnomalies}
\end{center}
\end{figure}

Alternatively, $t_{ijk}$ could take huge values if the coupling-full charges $q^{f-Z'}_{V/A}$ of a fermion $f$ is many orders of magnitude larger than $q^{\m-Z'}_{V/A}$.
Notice for example, that the coupling between $Z'$ and $\tau$ is the least constrained interaction in experiments if $Z'$ is purely leptonic force, which allows us to take $q^{\t-Z'}_{V/A}$ up to ${\cal O}(1)$\footnote{If $Z'$ also couples to quarks, tau neutrino experiments, such as DONuT experiment, may have sensitivity \cite{DONuT:2007bsg,Kling:2020iar}.}.

\begin{figure}[t]
\begin{center}
\epsfig{file=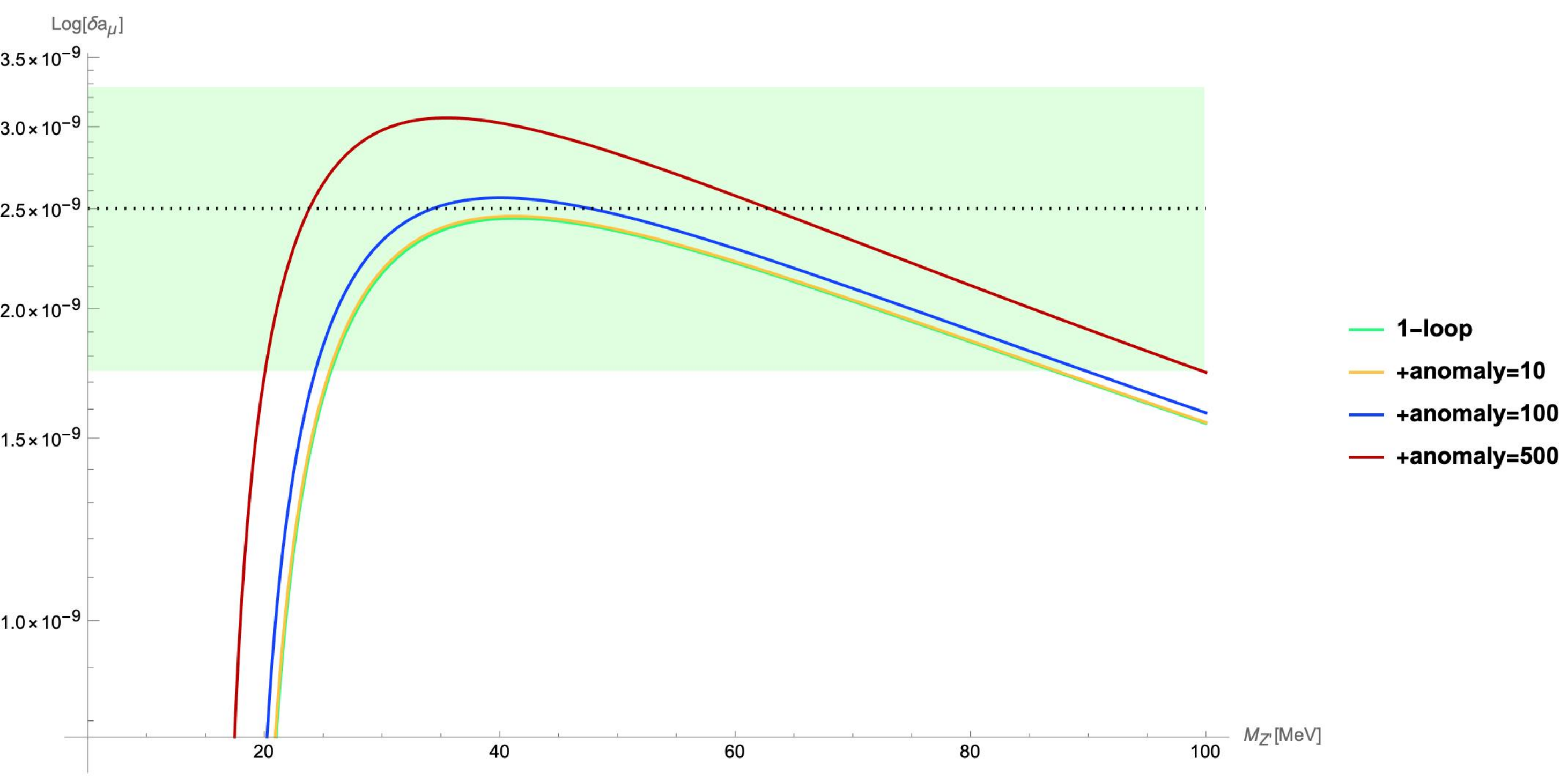,width=145mm}
\caption{In this plot, we add the one-loop and the anomalous contributions to the $g-2$ of the muon for masses $m_{Z'}=5-100$ MeV and fixed couplings
$g_{Z'}q_V^{\mu-Z'}=10\times g_{Z'} q_A^{\mu-Z'}=8\times 10^{-4}$.
We use for the anomaly $t_{ijk}=10,100,500$.}\label{Fig:figAnomalies+1loop}
\end{center}
\end{figure}

In fig \ref{Fig:fig1} we present two plots. The upper plot gives the one-loop contribution for different values of $g_{Z'}q_V^{\mu-Z'}=10\times g_{Z'} q_A^{\mu-Z'}=(4-10)\times 10^{-4}$.
The upper plot shows the contribution of the anomalous diagrams without the overall anomaly $t_{ijk}$ for similar values of
$g_{Z'}q_V^{\mu-Z'}$, $g_{Z'} q_A^{\mu-Z'}$.
For large values of the anomaly, the contribution of the anomalous diagram becomes leading and can absorb the full experimental discrepancy \eqref{Eq:exp}.
As an example, we fix the values
$g_{Z'}q_V^{\mu-Z'}=10\times g_{Z'} q_A^{\mu-Z'}=8\times 10^{-4}$ and we compare the one-loop contribution with the anomalous contribution for values of the anomalous trace $t_{ijk} = 10, 100, 500$ in fig \ref{Fig:figManyAnomalies}, and in fig \ref{Fig:figAnomalies+1loop} we add the one-loop and the anomalous contributions for the above values of the anomaly.

It is very interesting to note that for $m_{Z'}\sim 20$ MeV and $g_{Z'}q_V^{\mu-Z'}=10\times g_{Z'} q_A^{\mu-Z'}$ the one-loop contribution becomes very small due to the competing vectorial and axial coupling, and higher loop non-anomalous effects cannot contribute enough. However, if the $Z'$ is anomalous, with a large anomaly, the axionic and GCS terms take the lead, and they can reproduce the full discrepancy \eqref{Eq:exp}.

As a final but important remark,  we stress that the anomaly contribution is an effect that also reflects the UV properties of the theory.
Although its contribution to $g-2$ is of two-loop order, it is of a different nature than the other two-loop contributions. In particular, if the standard two-loop contribution is comparable to the one-loop contribution, this signals the breakdown of perturbation theory. If the anomalous contribution is comparable to the one-loop contribution, this is not a signal of the breakdown of perturbation theory, as the anomaly is one loop only.

\section{Leptophilic $Z'$ in D-brane realizations of the SM}\label{D-branes}

In this section, we shall discuss to what extent string theory models are possible where the $Z'$ couples only to leptons and, most importantly, to $\mu$ and $\tau$ but not the electron.

Anomalous $U(1)$'s are common in D-brane realizations of the SM,  \cite{Kiritsis:2003mc,rev1,rev2} and are the generic low-energy states with non-zero masses in perturbative string theory,
 as their masses are one-loop effects, \cite{AK}.
In this section, we discuss the possibility that the extension of the SM with one or more anomalous U(1)'s that couple to leptons (and specifically to the muon and tau and not to the electron) with a coupling of the order of $10^{-4}$, can be realized in semi-realistic D-brane configurations.

In this framework, SM particles are fluctuations of a ``local set of D-branes''. ``Local'' means a set of branes wrapping various cycles of the internal compact six-dimensional manifold and intersecting in an area in transverse space whose linear size is of the order of or smaller than the string length.
The SM particles, in this context,  are realized by the lightest fluctuations of open strings stretched between the local stack of D-branes, \cite{AKT,AIQU,AKRT, AD,ADKS, Anastasopoulos:2009mr,Anchordoqui:2021lmm,Anchordoqui:2022kuw}.
This, as a result, realizes the Standard model spectrum in terms of only bi-fundamental representations of the local D-brane gauge group\footnote{This is also the same requirement so that the SM spectrum is such that it can be coupled to any hidden sector in terms of bi-fundamental messengers. This is an important ingredient in models of emergent gravity \cite{Betzios:2020sro}, where axions \cite{Anastasopoulos:2018uyu}, graviphotons/dark-photons \cite{Betzios:2020sgd, Anastasopoulos:2020xgu}, neutrinos \cite{Anastasopoulos:2022sji} also emerge with special properties \cite{Anastasopoulos:2020gbu, Anastasopoulos:2021osp}.}.
In particular, the general rules can be summarised as follows:
\bi
\item[(a)] strings with both ends on a stack of $N$ parallel D-branes transform in the adjoin of $U(N)$ which splits in $SU(N)\times U(1)_N$ and

\item[(b)] strings stretched from a stack of $N$ to a stack of $M$ D-branes transform under $(N,+1;\bar M,-1)$, where $\pm 1$ are the charges under the abelian parts.
\ei
The SM's $SU(3)\times SU(2)\times Y$ is typically described by a stack of three, two, and some single D-branes.
Next, we would like to analyze different cases according to the number of stacks where the SM strings/particles are located:
\bi
\item Three stack models \cite{AD}.

If the SM particles are described by strings stretched between three different stacks of D-branes \cite{AD}, there is no linear combination between these three abelian factors that describes the lepton number. Therefore, a model which contains a leptophilic $Z'$ cannot be realized with three D-brane stacks.

\item Four stack models \cite{ADKS}.

In four stack models, the hypercharge and the Lepton number can be realized as a linear combination of the four abelian factors coming from each stack.
The different hypercharge and lepton number embeddings are given in \cite{ADKS}.
However, in all these cases, the hypercharge and the Lepton numbers are not orthogonal, allowing for large kinetic mixing and questioning the viability of such a  model at the phenomenological level.

We also searched cases where an ``approximate'' Lepton number, orthogonal to the hypercharge, is realized, where leptons have almost $\pm 1$ changes and the rest of the SM particles are almost zero. However, no such case was found.

\item Five stack models or more.

In this case, configurations where the hypercharge is a linear combination of four U(1)'s, and a universal Lepton number is coming from the fifth D-brane are possible \cite{Antoniadis:2021mqz}. Here, $Y$ and $L$ are orthogonal by construction.
A sixth D-brane is required to have a non-universal Lepton number where the muon and tau are charged, and the electron is not. The muon and tau are strings with one endpoint on the fifth D-brane (where the Lepton number lives), and the electron is a string with one endpoint on the sixth D-brane. The leptophilic U(1) coupling can take values of the order of $10^{-4}$ depending on the volumes that this D-brane wraps.

\ei

\section{Conclusions}

In this work, we have computed several contributions from an extra anomalous gauge
boson $Z'$ to the $g-2$ of the muon. We supplemented the one-loop contribution with additional contributions that are due to anomalous couplings of the Chern-Simons type. These higher-order contributions are unavoidable if the SM spectrum is anomalous concerning the U(1) of the $Z'$. Such (gauge) anomalies are cancelled by the effective generalized CS terms.
In the UV theory, the anomalies are cancelled by extra groups of fermions with masses well above the UV cutoff of the EFT.
Integrating out these anomaly cancelling fermions induces the anomaly cancelling generalized CS terms in the EFT.
We have shown that the anomalous contributions to $g-2$ are of two-loop order and therefore {\em generically} subleading to the one-loop (non-anomalous) contributions.
In this generic case, once a $Z'$ is discovered, measuring the subleading contributions provides a window into the anomalous massive spectrum of the theory.

However, depending on the $Z'$ charges of leptons, the one-loop contribution may be unnaturally small due to cancellations.
In such a case, the anomalous contributions may become comparable or dominant to the one-loop ones.
This does not signal the breakdown of perturbation theory as the anomalous vertex is one loop only and does not obtain higher loop corrections. In this second case, the $g-2$ excess could be explained in the region of parameter spaces $m_{Z'}\simeq 40$ MeV, and charges of the muon to the $Z'$,
$g_{Z'}(q_V^{\mu-Z'},q_A^{\mu-Z'})\simeq (10^{-3},10^{-4})$. The axial and vectorial contributions arise with negative relative signs, and
there are regions
of the parameter space where the anomalous higher-order contributions
do dominate.
If the overall anomaly traces are large, the anomalous contribution is leading, and it can even reproduce the whole discrepancy \eqref{Eq:exp}.
We also note that in such cases the cutoff of the theory is low enough and this implies that new charged fermions are within the reach of future colliders.

\vskip 1.5cm

\section*{Acknowledgements}\label{ACKNOWL}
\addcontentsline{toc}{section}{Acknowledgements}

\noindent We would like to thank C. Coriano, E. Niederwieser and P. Sphicas for discussions. Special thanks to S. Oribe for valuable discussions and checks on various parts of this paper. P.A. was supported by FWF Austrian Science Fund via the SAP P30531-N27.
The work of K.K. was supported in part by the Ministry of Education, Culture, Sports, Science, and Technology (MEXT)
of Japan, the Japan Society for the Promotion of Science (JSPS), the Grant-in-Aid for Scientific Research (C)
19H01899. This project has received support from the European Union's Horizon 2020 research and innovation programme under the Marie Sklodowska-Curie Grant Agreement No 860881-HIDDeN and the IN2P3 Master Project UCMN.

\newpage

\appendix
\renewcommand{\theequation}{\thesection.\arabic{equation}}
\addcontentsline{toc}{section}{Appendices}
\section*{APPENDIX}

{
\section{A UV completion of the effective model}\label{An UV completion}

In this appendix, we present a UV completion of the effective model presented in section~\ref{SM+U(1)+axion}.

We consider an anomaly-free model with the SM fermions (massless) and several extra massless fermions $\chi_i$. We also  have two Higgs fields, $H$ the doublet Higgs field of the SM and a extra scalar field $S$.
\bi

\item The SM fermions $\psi_{SM,i}$ and Higgs $H$ are generically  charged under $U(1)_A$ beyond the charges they gave under the SM gauge group.

\item The extra fermions $\chi_i$ are charged under the $U(1)_Y\times U(1)_A\times SU(2)_W$ .

\item The extra scalar $S$ is charged only under $U(1)_A$, and it has Yukawa couplings only with the extra fermions $\chi_i$.
\ei
The action schematically is
\bea
{\cal L}={\cal L}_{SM}-{1\over 4} F^2_A+\bar \chi_i \gamma_\m D^\mu \chi_i + |DS|^2 + \lambda_S (|S|^2-V^2)^2 + Y_{ij} \bar \chi_i \chi_j S +c.c.
\label{UV action}
\eea
where ${\cal L}_{SM}$ is the action of the SM augmented by the minimal couplings of the SM fields to $U(1)_A$.

Anomaly conditions require that the contribution to the anomalies from all triangle diagrams (SM plus the extra fermions) cancel out.

We assume that $V\gg v_H$, where $v_H$ is the SM HIggs vev, and there is very small mixing between $H$ and $S$.
 
\bi

\item Upon $S$-induced symmetry breaking, the additional U(1)$_A$  gauge field ($A^\m$) becomes massive with mass $M= g_A q^{S-A} V$, where $q^{S-A}$ is the charge of $S$ under $A^\m$.

\item $S$ becomes massive with $m_S=\sqrt{2 \lambda_S} V$.

\item The fermions $\chi_i$ become massive, forming in pairs Dirac  fermions, with  masses $Y_{ij}V$, where $Y_{ij}$ are the Yukawa couplings in (\ref{UV action}).

\ei
We assume that the S-mass as well as the extra fermion masses are much larger than the mass $M$ of $A_{\m}$ 
\be
M\ll \Lambda_{S,\chi}\equiv  min\{\sqrt{2 \lambda_S } V,  Y_{ij}V\}
\ee
and we integrate out the $S$ field as well as the extra massive fermions $\chi_i$.
This defines our effective field theory. It contains the Standard model fields, the $U(1)_A$ gauge boson , and the phase of $S$ that appears as a low-energy axion field.
(see a schematic representation of our toy-model in fig. \ref{Fig:toymodel}).

\begin{figure}[t]
\begin{center}
\includegraphics[width=150mm]{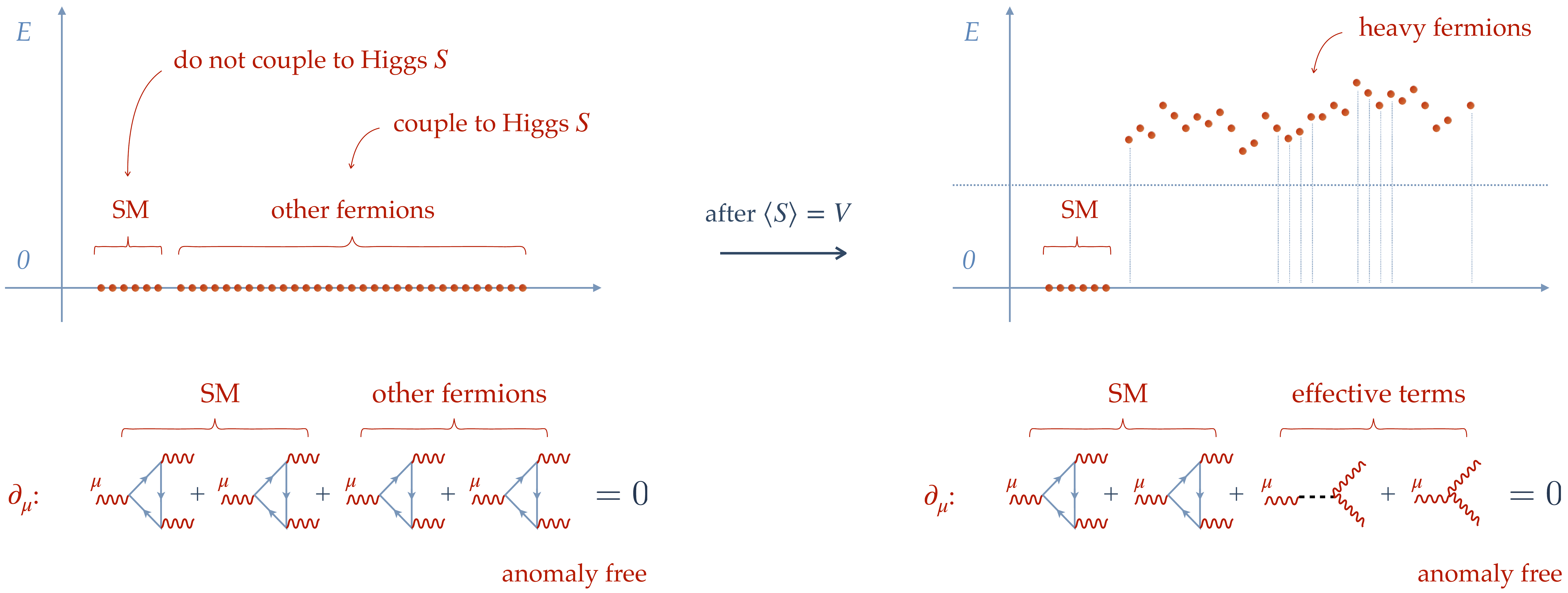}
\caption{A schematic representation of the toy-model. }
\label{Fig:toymodel}
\end{center}
\end{figure}

The theory described above is rather general. In many cases, the subset of fermion $\chi_i$ that become heavy and are integrated out, may have non-zero anomaly contributions
to the symmetries of the effective field theory, \cite{ABDK}.These contributions must be visible in the effective theory and indeed they appear from the axionic, and GCS effective terms \cite{ABDK, PA, Anastasopoulos:2008jt}.

\subsection{Obtaining a low mass $Z'$ (toy) model}\label{A low mass $Z'$ (toy) model}

As  mentioned above, some of the parameters of our fundamental theory are
\bi
\item the large vev of the Higgs field $S$ which we denote by $\langle S\rangle =V$,

\item the parameter $\lambda_S$ which also controls  the mass $m_{S}=\sqrt{2 \lambda_S} V$ of the heavy scalar $S$,

\item the Yukawa couplings $Y_{ij}$  responsible for giving heavy masses to all extra fermions $\chi_i$, and

\item the mass of $A^\m$, $m_{A}=g_A q^{S-A}V$.

\ei

Next, we assign some values to the parameters above.

We assume that the Yukawa couplings $Y_{ij}$ and the coupling $\lambda_S$
are in the  perturbation regime, $Y_{ij},\lambda_S \sim 10^{-1}$.
For the masses of the heavy fermions and the extra scalar to be above the 40 TeV, we obtain a value
\bea
V\sim 400\textrm{TeV}~~~~\to ~~~~ m_\textrm{heavy fermions},
m_{S}\sim 40\textrm{TeV}
\eea
With this value of the vev of $S$, and a coupling $g_{Z'}\sim 10^{-4}$, we obtain
\bea
g_{Z'} q^{S-Z'} V \sim 10-100\textrm{MeV}
~~~~\to ~~~~ q^{S-Z'}\sim 2.5\times 10^{-3} -2.5\times   10^{-2}
\eea
where $q^{S-Z'}$ is the charge of the scalar $S$ under the $A_{\m}$ vector field.

Finally, we would like to make some comments about the value of the tree-level anomaly of the effective theory.
In our effective theory, among the  anomaly-related contributions to $g-2$ , the dominant one is originating from ${\cal A}[Z',\g]$ \eqref{diagram for g-2 Z' A}.
This contributions is proportional to the anomaly trace $t_{Z'\g\g}$. 
Since we assume in the main text that the only SM fermions charged under the anomalous $Z'$ are the muon and tau leptons, we can evaluate
\bea
t_{Z'\g\g}= \sum_{f=\m,\t} q_A^{f-Z'} q_V^{f-\g}q_V^{f-\g} = q_A^{\m-Z'}+q_A^{\t-Z'}
\eea
since $q_A^{f-\gamma}=0$ for all SM fermions and $q_V^{\m/\t-\gamma}=-1$.
Assuming also that $q_A^{\t-Z'}\gg q_A^{\m-Z'}$, the axial charge of the tau lepton controls the anomaly.
Finally, imposing the unitarity bound $g_{Z'} q^{\tau-Z'}< {\cal O}(1)$, and  $g_{Z'}\sim 10^{-4}$,  we obtain
\bea
t_{Z'\g\g}\sim q_A^{\t-Z'}< 10^4\;.
\eea
This  allows for large anomaly values, that  control the anomalous contribution of the $g-2$ of the muon.
In the main part of the paper, we shall  explore the  range
\bea
50<t_{Z'\g\g}<500
\label{anomaly bounds}
\eea

}

\section{Feynman rules}\label{Feynman rules}

These are the Feynman rules needed for our computation:
\bi
\item Propagators (unitary gauge)
\bea
&&\g ~~~:~~~~~~~~
D_{\g}^{\m\n} (k)= {- i g^{\m\n} \over k^2}
\label{Eq:Dphoton}
\\
&&Z/Z' ~:~~~~~~~
D_{\tilde A}^{\m\n} (k)= { i \over k^2-m_{\tilde A}^2} \left(g^{\m\n} -{k^\m k^\n \over m_{\tilde A}^2} \right)
\label{Eq:DZZ'}
\eea
The fermions and bosons (Higgs) in the broken phase have propagators
\bea
&&\textrm{scalar} ~:~~~~~~
\D (p)= {i\over  p^2-m^2_s} \\
&&\textrm{lepton} ~:~~~~~
S (p)= {i\over \sla p-m_l}
\label{Eq:S}
\eea
\item Minimal couplings
\bea
\ba{llllllllll}
\textrm{lepton $\ell$ with photon}&&\to && - i g_{EM}  q^{\ell-\g}_V\g_\m\\
\textrm{lepton $\ell$ with $Z$}&& \to && -i g_{Z}  q^{\ell-Z}\g_\m=-i g_Z \left(q^{\ell-Z}_V \mathbb{I} - q^{\ell-Z}_A\g_5\right)\g_\m\\
\textrm{lepton $\ell$ with $Z'$}&& \to && -i g_{Z'}  q^{\ell-Z'}\g_\m=-i g_{Z'} \left(q^{\ell-Z'}_V \mathbb{I} - q^{\ell-Z'}_A\g_5\right)\g_\m\ea
\label{couplings to leptons}\eea
where $g_{EM}$, $g_Z=g_2/2\cos \theta_W$, $ g_{Z'}$ the coupling constants of the electromagnetism the $Z$ and the $Z'$ (see \eqref{couplings gauge fields to leptons}).
\ei

\subsection{SM leptons and gauge fields in the broken (photon) basis}

Applying the rotation matrix $O$ to the SM part of \eqref{Ltotal} and we collect the couplings between leptons and gauge bosons (in the broken phase)\footnote{We present only couplings to the leptons since this is the relevant part of \eqref{Ltotal} for the $g-2$ of the muon.}
\bea
{\cal L}_{leptons} &=& e \bar e_i \g_\m e_i A^\m_{\g}
- {g_2\over 2\sqrt{2}} W^+_\m \bar e_j \g^\m (1-\g_5) \n_i {\cal U}^\n_{ji}
- {g_2\over 2\sqrt{2}} W^-_\m \bar \n_j \g^\m (1-\g_5) e_i {\cal U}^{\n\dagger}_{ji} \nn\\
&&
- {g_2\over 2\cos \theta_W} Z_\m \bar e_i (q_V^{e-Z} \g^\m - q_A^{e-Z} \g^\m \g_5 ) e_i
- {g_2\over 2\cos \theta_W} Z_\m \bar \n_i (q_V^{\n-Z} \g^\m - q_A^{\n-Z} \g^\m \g_5 ) \n_i \nn\\
&&
- g_{Z'} Z'_{\m} \bar e_i (q_V^{e-Z'} \g^\m - q_A^{e-Z'} \g^\m \g_5 ) e_i
- g_{Z'} Z'_{\m} \bar \n_i (q_V^{\n-Z'} \g^\m - q_A^{\n-Z'} \g^\m \g_5 ) \n_i
\label{couplings gauge fields to leptons}\eea
where $ij=1,2,3$ for the different families of the SM and ${\cal U}^\n_{ji}$ the mixing matrix of the leptons.

\section{The anomalous coupling}

As we have already mentioned, the anomalous coupling splits into two parts
\bea
\G_{\m\n\r}^{ijk}|_\textrm{total}&=&\G_{\m\n\r}^{ijk}|_\textrm{one-loop} + \G_{\m\n\r}^{ijk}|_\textrm{axion} + \G_{\m\n\r}^{ijk}|_\textrm{GCS}\label{total anomalous coupling}\\
&=& \Big(\G_{\m\n\r}^{ijk}|^\textrm{C=even}_\textrm{one-loop}\Big)
+\Big(\G_{\m\n\r}^{ijk}|^\textrm{C=odd}_\textrm{one-loop}+ \G_{\m\n\r}^{ijk}|_\textrm{axion} + \G_{\m\n\r}^{ijk}|_\textrm{GCS}\Big)\nn
\eea
where the C even part identically vanishes due to Furry's theorem. Therefore, in diagrams
\bea
\scalebox{0.7}{\begin{tikzpicture}[thick,baseline={-0.1cm}]
  \begin{feynman}[every blob={/tikz/fill=gray!30,/tikz/inner sep=2pt}]
    \vertex[blob] (f1) at (0,0) {$~~~~{\cal A}~~~~$};
      \vertex (a) at (-2.5,0) {\(A_i^\m\)};
      \vertex (b) at (2,2) {\(A_k^\r\)};
      \vertex (c) at (2,-2) {\(A_j^\n\)};
\diagram* {
      (a) -- [boson, edge label=\(k_3\)]
      (f1) -- [boson, edge label=\(k_2\)] (b),
      (f1) -- [boson, edge label'=\(k_1\)] (c)};
  \end{feynman}
\end{tikzpicture}}
&=&
\left(
\scalebox{0.7}{\begin{tikzpicture}[thick,baseline={-0.1cm}]
  \begin{feynman}
    \vertex (a) {\(A_i^\m\)};
    \vertex [right=of a] (f1);
    \vertex [above right=of f1] (f2);
    \vertex [below right=of f1] (f3);
    \vertex [right=of f2] (b) {\(A_k^\r\)};
    \vertex [right=of f3] (c) {\(A_j^\n\)};
    \diagram* {
      (a) -- [boson, edge label=\(k_3\)]
      (f1) -- [fermion, edge label'=\(p-k_1\)] (f3),
      (f3) -- [fermion, edge label'=\(p\)] (f2),
      (f2) -- [fermion, edge label'=\(p+k_2\)] (f1),
      (f2) -- [boson, edge label=\(k_2\)] (b),
      (f3) -- [boson, edge label=\(k_1\)] (c)};
  \end{feynman}
\end{tikzpicture}}
%
+
%
\scalebox{0.7}{\begin{tikzpicture}[thick,baseline={-0.1cm}]
  \begin{feynman}
    \vertex (a) {\(A_i^\m\)};
    \vertex [right=of a] (f1);
    \vertex [above right=of f1] (f2);
    \vertex [below right=of f1] (f3);
    \vertex [right=of f2] (b) {\(A_k^\r\)};
    \vertex [right=of f3] (c) {\(A_j^\n\)};
    \filldraw[black] (3,-.8) circle (0pt) node[anchor=west]{$k_1$};
    \filldraw[black] (3,.8) circle (0pt) node[anchor=west]{$k_2$};
        \diagram* {
      (a) -- [boson, edge label=\(k_3\)]
      (f1) -- [fermion, edge label'=\(p-k_1\)] (f3),
      (f3) -- [fermion, edge label'=\(p\)] (f2),
      (f2) -- [fermion, edge label'=\(p+k_2\)] (f1),
      (f2) -- [boson
      ] (c),
      (f3) -- [boson
      ] (b)};
  \end{feynman}
\end{tikzpicture}}\right)_\textrm{CP=odd}\nn\\
&&
+
\scalebox{0.7}{\begin{tikzpicture}[thick,baseline={-0.1cm}]
  \begin{feynman}
    \vertex (a) {\(A_i^\m\)};
    \vertex [right=of a] (f1);
    \vertex [right=of f1] (f2);
    \vertex [above right=of f2] (b) {\(A_k^\r\)};
    \vertex [below right=of f2] (c)
    {\(A_j^\n\)};
\diagram* {
      (a) -- [boson, edge label=\(k_3\)]
      (f1) -- [scalar, edge label=\(a\)] (f2),
      (f2) -- [boson, edge label=\(k_2\)] (b),
      (f2) -- [boson, edge label'=\(k_1\)] (c)};
  \end{feynman}
\end{tikzpicture}}
+
\scalebox{0.7}{\begin{tikzpicture}[thick,baseline={-0.1cm}]
  \begin{feynman}
    \vertex (a) {\(A_i^\m\)};
    \vertex [right=of a] (f1);
    \vertex [above right=of f1] (b) {\(A_k^\r\)};
    \vertex [below right=of f1] (c)
    {\(A_j^\n\)};
\diagram* {
      (a) -- [boson, edge label=\(k_3\)]
    (f1) -- [boson, edge label=\(k_2\)] (b),
      (f1) -- [boson, edge label'=\(k_1\)] (c)};
  \end{feynman}
\end{tikzpicture}}
+...
\label{anomalousdiagramsFIG}\eea
where the ellipsis indicates that we should take all permutations of the axionic propagator with all other legs.
The parametrization has been analytically evaluated in \cite{ABDK}, and here we present the results
\bea
&&\G^{ijk}_{\m\n\r}|_\textrm{one-loop}^\textrm{CP=odd}=
g_i g_j g_k t^{ijk} \Big( \e_{\m\n\r\s} (A_1  k_2^\s
+A_2 k_1^\s )\label{one-loop-CPodd_tensor}\\
&&
~~~~~~~~~~~~~~~~~~~~~~~+
(B_1 k_{2\n} +B_2 k_{1\n} )
\e_{\m\r\s\t} k_2^\s k_1^\t
+
(B_3 k_{2\r} + B_4 k_{1\r}) \e_{\m\n\s\t} k_2^\s k_1^\t
\Big)  \nn\\
&&\G^{ijk}_{\m\n\r}|_\textrm{axion}= -g_i g_j g_k\Big(M^i_I C^{jk}_I {k_{3\m} \over
k_3^2} \e_{\n\r\s\t} k_2^\s k_1^\t\nn\\
&& ~~~~~~~~~~~~~~~~~~~~~~~~~~~~
+M^j_I C^{ki}_I {k_{1\n} \over k_1^2} \e_{\r\m\t\s}
k_2^\s k_3^\t 
+M^k_I C^{ij}_I {k_{2\r} \over k_2^2} \e_{\m\n\t\s}
k_3^\s k_1^\t\Big) ~~~~~~~~\label{axion_tensor}\\
&&\G^{ijk}_{\m\n\r}|_\textrm{GCS}= -g_i g_j g_k\Big(E^{ij,k} \e_{\m\n\r\s} k_2^\s
+E^{jk,i} \e_{\n\r\m\s} k_3^\s +E^{ki,j} \e_{\r\m\n\s} k_1^\s\Big)
\label{GCS_tensor}\eea
where the anomaly coefficient $t^{ijk}$ contains the trace anomaly and $g_{i/j/k}$ the minimal couplings between the fermions in the loop and the gauge fields $A^\m_{i/j/k}$.
The $B_1,B_2,B_3,B_4$ are finite coefficients (functions of $k_1,k_2$)
\bea
&& B_1(k_1,k_2)=-B_4(k_2,k_1)=-{i\over 8\p^2} \int^1_0 d\a \int^{1-\a}_0d\b {2\a\b \over \a k^2_2 + \b k^2_1 - (\a k_2-\b k_1)^2-{m_f^2}}~~~~~~~~~\label{B1relations}\\
&& B_2(k_1,k_2)=-B_3(k_2,k_1)=-{i\over 8\p^2} \int^1_0 d\a \int^{1-\a}_0 d\b {2\b(1-\b) \over \a k^2_2 + \b k^2_1 - (\a k_2-\b k_1)^2 -{m_f^2}}~~~~~~\label{B2relations}\eea
Note that the above coefficients (in particular \eqref{B2relations}) contain an IR divergence in the massless $m_f\to 0$ limit.

We should mention that the (scheme-dependant) coefficients $A_1, A_2$ can be eliminated in terms of the finite and unambiguous coefficients $B_i$'s, by using Bose symmetry and $C_A$ the standard anomaly coefficient \cite{ABDK}.
\bea
&& A_1=-{C_A\over 3} -k_1\cdot k_2 B_1-k_1^2 B_2\\
&& A_2=+{C_A\over 3}-k_2^2 B_3-k_1\cdot k_2 B_4\\
&& A_1-A_2= {C_A\over 3}
\eea
Requiring that the Ward identities be satisfied,  we obtain the anomaly conditions for the most generic case to be \cite{ABDK}
\bea
\left.
\ba{lllll}&k_1^\n\G^{ijk}_{\m\n\r}|_\textrm{total}=0\\
&k_2^\r\G^{ijk}_{\m\n\r}|_\textrm{total}=0\\
&k_3^\m\G^{ijk}_{\m\n\r}|_\textrm{total}=0
\ea
\right\}~~\to ~~
\ba{rlllll}
& C_A~t_{ijk} &=& M_I^k C^{ij}_I + M_I^j C^{ki}_I + M_I^i C^{jk}_I\\
& 3 E^{ij,k} &=& M_I^i C^{jk}_I - M_I^j C^{ki}_I \\
& 3 E^{jk,i} &=& M_I^j C^{ki}_I - M_I^k C^{ij}_I \\
& 3 E^{ki,j} &=& M_I^k C^{ij}_I - M_I^i C^{jk}_I
\ea
\label{genericAnomalousConditionsAGAIN}
\eea
where summation over $I$ is understood. In these conditions, we have considered the coefficients' structure. Therefore, the symmetric combination of the $M_I^{(i}C_I^{jk)}$ cancels the democratically distributed anomaly, and the antisymmetric part $M_I^{[i}C_I^{j]k}$ cancels the GCS terms.

\subsection*{The anomalous coupling in our case}

In this section, we use the anomalous conditions found in \eqref{anomalyConditionAY} to give the form of the different couplings (we use that $k_3=-k_1-k_2$)
\bi
\item The $AYY$ anomalous coupling

Using the anomaly conditions \eqref{anomalyConditionAY}
we obtain the anomaly relations
\bea
E_{AY,Y}={C_A} t_{AYY}~,~~~~
M C_{YY}=-2C_A t_{AYY}
\eea
and inserting them in \eqref{one-loop-CPodd_tensor}, \eqref{axion_tensor},
\eqref{GCS_tensor} and using \eqref{B1relations}, \eqref{B2relations}, we obtain the contributions from the triangle, the axionic and GCS terms as
\bea
&&\G^{AYY}_{\m\n\r}|_\textrm{one-loop}^\textrm{CP=odd}=
\G^{AYY}_{\m\n\r}|_\textrm{one-loop}^\textrm{mass-independent}+
\G^{AYY}_{\m\n\r}|_\textrm{one-loop}^\textrm{mass-dependent}
\label{one-loop-CPodd_tensorAYY}\\
&&\G^{AYY}_{\m\n\r}|_\textrm{one-loop}^\textrm{mass-independent}=
-g_Ag_Y^2 t_{AYY}
{C_A\over 3} \e_{\m\n\r\s} (k_2^\s-k_1^\s)  \label{one-loop-CPodd_tensorAYY_NoMass}\\
&&\G^{AYY}_{\m\n\r}|_\textrm{one-loop}^\textrm{mass-dependent}=
g_Ag_Y^2 \sum_f q^{f-A}q^{f-Y}q^{f-Y}\Big( -\Big(k_1\cdot k_2 B_1^f + k_1^2 B^f_2\Big) \e_{\m\n\r\s} k_2^\s\nn\\
&& ~~~~~~~~~~~~~~~~~~~~~~~~~~~~~~~~~~~~~~~~~~~~~~~~~~~~ -\Big(k_2^2 B^f_3+k_1\cdot k_2 B^f_4 \Big) \e_{\m\n\r\s}  k_1^\s ~~~~~~\nn\\
&& ~~~~~~~~~~~~~~~~~~~~~~~~~~~~~~~~~~~~~~~~~~~~~~~~~~~~ +(B^f_1 k_{2\n} +B_2^f k_{1\n} )\e_{\m\r\s\t} k_2^\s k_1^\t\nn\\
&& ~~~~~~~~~~~~~~~~~~~~~~~~~~~~~~~~~~~~~~~~~~~~~~~~~~~~ +(B_3^f k_{2\r} + B_4^f k_{1\r}) \e_{\m\n\s\t} k_2^\s k_1^\t \Big)  \label{one-loop-CPodd_tensorAYY_mass}\\
&&\G^{AYY}_{\m\n\r}|_\textrm{axion}=
2g_Ag_Y^2t_{AYY} C_A ~{k_{3\m} \over
k_3^2} \e_{\n\r\s\t} k_2^\s k_1^\t ~~~~~~~~\label{axion_tensorAYY}\\
&&\G^{AYY}_{\m\n\r}|_\textrm{GCS}= -g_Ag_Y^2 t_{AYY} {C_A} \e_{\m\n\r\s} (k_2^\s - k_1^\s)
\label{GCS_tensorAYY}\eea
%
%
%
%
%
%
where we use for compactness $B_i^f=B_i^f[k_1,k_2]$ for $i=1,..,4$, for each different fermion $f$ in the loop.

On the other hand, the axionic and GCS terms contribute with the full anomaly $t^{AYY}$ in both the unbroken and broken phases.

\item The $YAA$ anomalous coupling

Similarly to the above, for the case of one anomalous and two hypercharges, the anomaly conditions give
\bea
E_{YA,A}={C_A} t_{YAA}~,~~~~
M C_{YA}=-2{C_A} t_{YAA}
\eea
and the $YAA$ coupling is given by
\bea
&&\G^{YAA}_{\m\n\r}|_\textrm{one-loop}^\textrm{CP=odd}=
\G^{YAA}_{\m\n\r}|_\textrm{one-loop}^\textrm{mass-independent}+
\G^{YAA}_{\m\n\r}|_\textrm{one-loop}^\textrm{mass-dependent}
\label{one-loop-CPodd_tensorYAA}\\
&&\G^{YAA}_{\m\n\r}|_\textrm{one-loop}^\textrm{mass-independent}=
-g_Yg_A^2t_{YAA}{C_A\over 3} \e_{\m\n\r\s} (k_2^\s-k_1^\s)  \label{one-loop-CPodd_tensorYAA_NoMass}\\
&&\G^{YAA}_{\m\n\r}|_\textrm{one-loop}^\textrm{mass-dependent}=g_Yg_A^2
\sum_f q^{f-Y}q^{f-A}q^{f-A}\Big( -\Big(k_1\cdot k_2 B_1^f + k_1^2 B^f_2\Big) \e_{\m\n\r\s} k_2^\s\nn\\
&& ~~~~~~~~~~~~~~~~~~~~~~~~~~~~~~~~~~~~~~~~~~~~~~~~~~~~ -\Big(k_2^2 B^f_3+k_1\cdot k_2 B^f_4 \Big) \e_{\m\n\r\s}  k_1^\s ~~~~~~\nn\\
&& ~~~~~~~~~~~~~~~~~~~~~~~~~~~~~~~~~~~~~~~~~~~~~~~~~~~~ +(B^f_1 k_{2\n} +B_2^f k_{1\n} )\e_{\m\r\s\t} k_2^\s k_1^\t\nn\\
&& ~~~~~~~~~~~~~~~~~~~~~~~~~~~~~~~~~~~~~~~~~~~~~~~~~~~~ +(B_3^f k_{2\r} + B_4^f k_{1\r}) \e_{\m\n\s\t} k_2^\s k_1^\t \Big)  \label{one-loop-CPodd_tensorYAA_mass}\\
&&\G^{YAA}_{\m\n\r}|_\textrm{axion}=
2g_Yg_A^2 t_{YAA} {C_A} ~\Big(
{k_{1\n} \over k_1^2} \e_{\r\m\t\s} + {k_{2\r} \over k_2^2} \e_{\m\n\t\s} \Big) k_2^\s k_1^\t ~~~~~~~~\label{axion_tensorYAA}\\
&&\G^{YAA}_{\m\n\r}|_\textrm{GCS}= -g_Yg_A^2 t_{YAA} {C_A} \e_{\m\n\r\s} (k_2^\s - k_1^\s)
\label{GCS_tensorYAA}\eea
%
%
%
%
with the same definitions of $B_i^f$.
\ei
From the above expressions, we can evaluate the anomalous couplings in the broken phase by the rotation \eqref{YW3A to gamma Z Z' basis}. That will just modify the anomaly coefficients.

\section{Extracting the contributions for the $g-2$ from the anomalous diagrams}\label{Extracting g-2 and EDM}

In this section, we provide some technical details of the extraction of the contribution of the anomalous diagrams to $g-2$ of the leptons.

The diagrams (\ref{diagram for g-2 Z' A}-\ref{diagram for g-2 Z Z}) are lengthy and complicated.
Here, we intend to search for the specific terms within these diagrams that correspond to $g-2$, without evaluating the full diagrams.

The couplings $q^\ell_{\g/Z/Z'}$ should be taken in accordance with the propagating gauge fields in the loop.
Note that $q^\ell_{Z/Z'}$ contains $P_L/P_R$ projection operators. We should also include the diagram with exchanged propagating gauge fields in the loop.

The steps for this lengthy procedure are the following:
\bi

\item Before we start manipulating the diagrams, we split the anomalous coupling (triangle, axion and GCS) into three parts:
\bi
\item[a.] the contribution of the axion and GCS terms,
\item[b.] the $m_f$ independent triangle part and
\item[c.] the $m_f$ dependent triangle part.
\ei
The first two depend on the full anomaly (and they are absent in the non-anomalous case); however, the last one is proportional to the charges of each massive fermion in the loop and is also present in the non-anomalous case.

\item Next, we expand the different diagrams (\ref{diagram for g-2 Z' A}-\ref{diagram for g-2 Z Z}). The result is several terms with different spinorial lines. We classify them according to the number of gamma matrices $\g_\m$ between the spinors $\bar u(p')... u(p)$.

For this purpose, we use {\it Mathematica} and the package {\it FeynCalc}.

\item Next, we use the Feynman parametrization that leads to integration over loop momentum $k$.

Note that we have three propagating gauge fields in the loop. However, within the anomalous couplings, there are extra axionic propagations and the coefficients $B_1$,..,$B_4$ \eqref{B1relations}, \eqref{B2relations} where the loop momentum appears in the denominator. We consider these coefficients $B_1$,..,$B_4$ as effective propagators, and we perform the Feynman parametrization for three or (effectively) four propagating fields.

Schematically, we have (propagators are provided in section \ref{Feynman rules} and $\Delta_{axion}^{-1}(k)=-ik^2$)
\bea
D_{\g,Z,Z'}[k]~S_{fermion}[p-k]~D_{\g,Z,Z'}[-k-q]\times(\Delta_{axion}, B_1, B_2, B_3, B_4)
\eea
leading to different denominators. For example, the Feynman parametrization for one of them will be
\bea
&&D_{\g}[k]~S_{fermion}[p-k]~D_{Z'}[-k-q]~B_1[q,k] \label{second parametrization}\\
&&~~~~~\sim \Big((k^2)((p-k)^2-m_l^2)((k+q)^2-m^2_{Z'}) (\a k^2 + \b q^2 - (\a k-\b q)^2-{m^2_f}) \Big)^{-1}~~~~~
\nn\\
&&~~~~~=\int dxdydzdw~ 3!~\d(x+y+z+w-1)
\nn\\
&&~~~~~~~~\times
\Big(x k^2+y((p-k)^2-m_l^2)+z((k+q)^2-m^2_{Z'})\nn\\
&&~~~~~~~~~~~~~~~~~~~~
~~~~~~~~~~~~~~~~~~~~~~~~
+w (\a k^2 + \b q^2 - (\a k-\b q)^2-{m^2_f}) \Big)^{-4}~~~~\nn
\eea
leading to
\bea
&&\textrm{momentum shift :}~~~ k\to K-(abw+z) q+y p\\
&&\textrm{denominator ~~~~~:}~~~ (K^2-(y^2 m^2+w m^2+ z m_{Z'}^2))^4
\eea
after applying equations of motion for the external fields
\bea
q^2=0~,~~~~ p^2=p'^2=m_f^2~,~~~~ q\cdot p=0
\label{eom}
\eea

The above is applied five times for each of the five different diagrams (\ref{diagram for g-2 Z' A}-\ref{diagram for g-2 Z Z}).

\item After Feynman parametrization, we drop odd terms of the shifted loop momentum $K$ from the numerator, and we apply the following identities
\bea
&&\s^{\m\n}={i\over 2}(\g^\m \g^\n -\g^\n
\g^\m)
~~\textrm{and}~~
\s^{\m\n}\g_5={i\over 2}\s_{\r\s} \e^{\m\n\r\s}\\
&&~~~~~
\g^\m\g^\n=g^{\m\n}-i\s^{\m\n}\\
&&~~~~~
\g^\m\g^\n\e_{\m\n\r\s}=-i \s^{\m\n}\e_{\m\n\r\s}
=-2 \s_{\r\s}\g_5\\
&&~~~~~
\g^\m\g^\n\g_5
\e_{\m\n\r\s}
=-2 \s_{\r\s}\g_5\g_5
=-2 \s_{\r\s}
\eea
as well as the Gordon identity
\bea
\bar u(p') \g^\m u(p) = \bar u(p') \left( {(p'+p)^\m\over 2 m} +
{i \s^{\m\n} q_\n\over 2 m}\right) u(p)\eea
in order to extract the contributions to $g-2$ $F_2(q)$ \eqref{g-2 and EDM}.


\item Next, we integrate over loop momenta $K$. Since the divergent part is cancelled, we use the standard integrals
\bea
&&\int {d^4K\over (2\p)^4}
{1\over (K^2-M^2)^4}
=
{i\over (4\p)^2}{\G[2]\over \G[4]} (M^2)^{-2}\nn\\
&&\int {d^4K\over (2\p)^4}
{K^\m K^\n\over (K^2-M^2)^4}
=
-{i\over (4\p)^2}{g^{\m\n}\over 2}{\G[1]\over \G[4]} (M^2)^{-1}
\eea

\item The last step is to integrate over five different Feynman parameters. Two parameters, $a,b$, are coming from the coefficients $B_1,...,B_4$ and three, $x,y,z$, from the second parametrization \eqref{second parametrization}.

Integrals over $a,b$ can be performed using {\it Mathematica}.

\ei
The final results for specific values for the masses and couplings of the contributing fields are given in section \ref{g-2 analysis and results}.

\begin{figure}[t]
\begin{center}
\raisebox{25mm}[0pt][0pt]{${\cal A}[Z',\g]$} ~~~ \epsfig{file=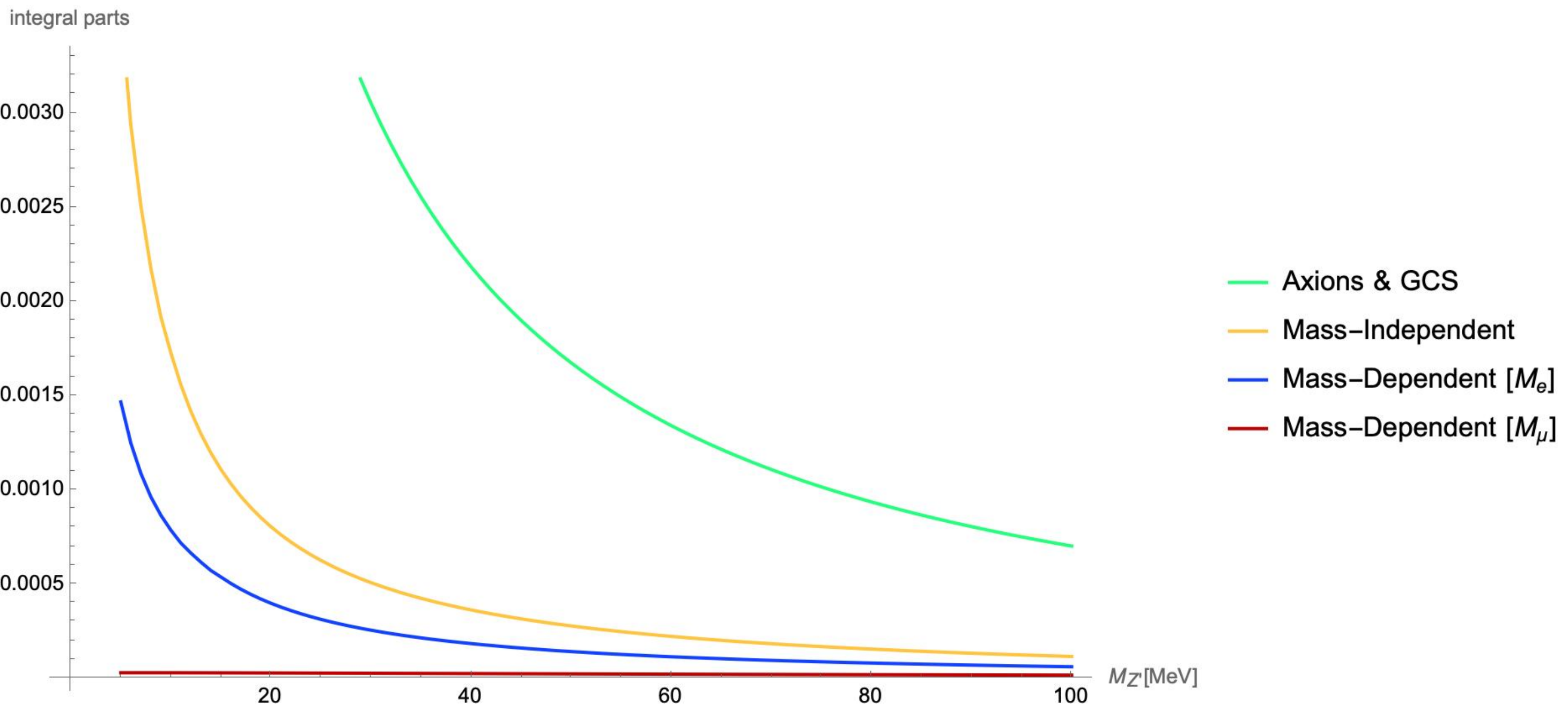,width=125mm}\\
\raisebox{25mm}[0pt][0pt]{${\cal A}[Z',Z]$} ~~~ \epsfig{file=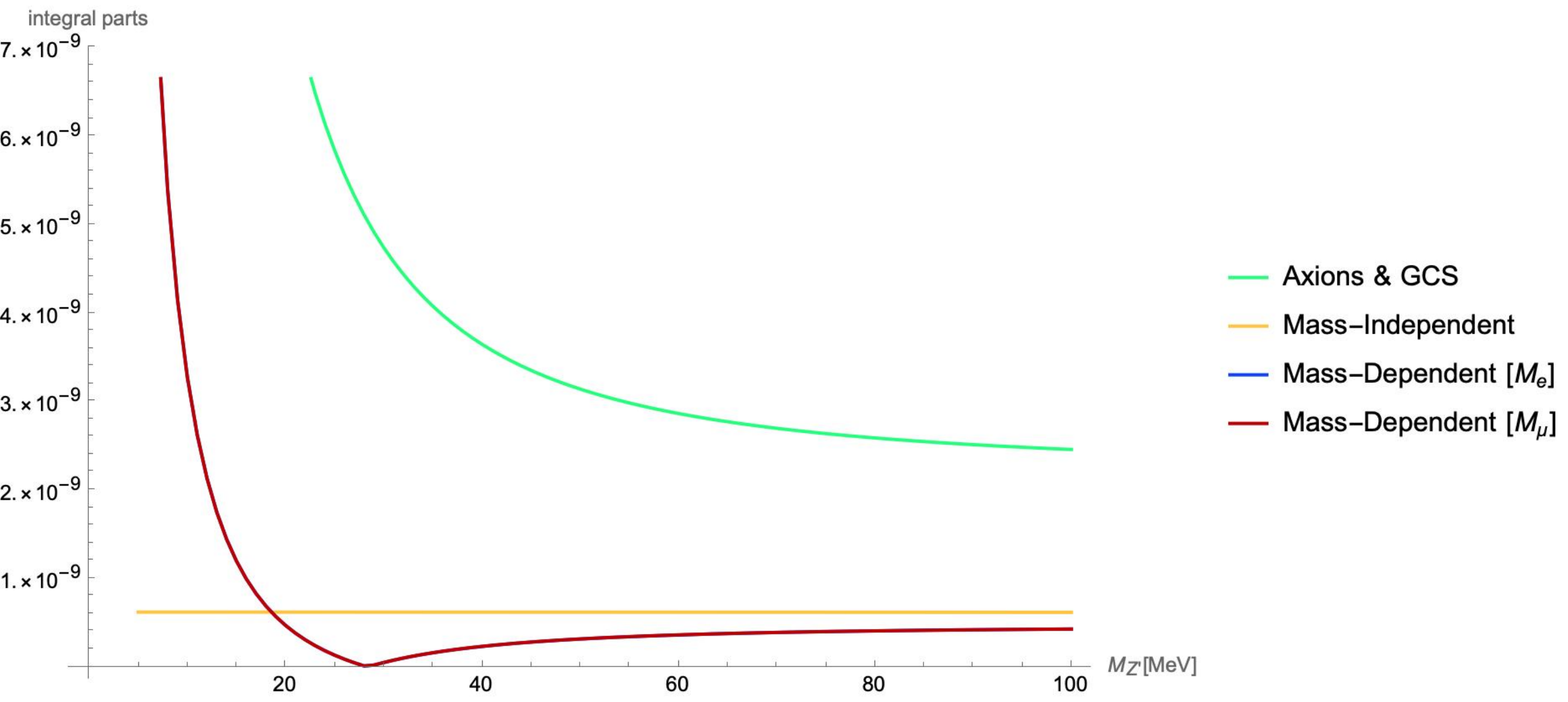,width=125mm}\\
\raisebox{25mm}[0pt][0pt]{${\cal A}[Z',Z']$} ~~~ \epsfig{file=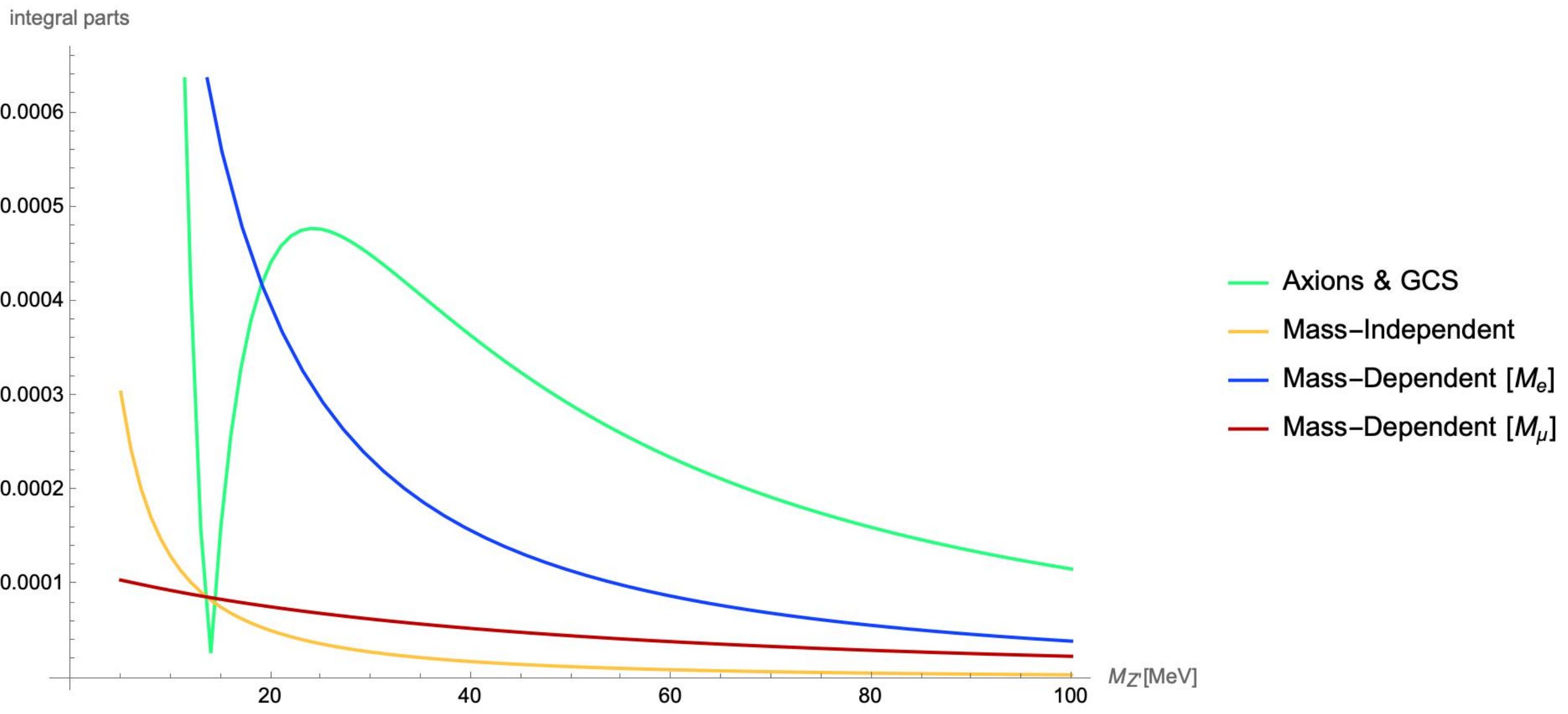,width=125mm}
\caption{In these plot we compare the absolute value of the contribution coming from the amplitudes ${\cal A}[Z',\g]$, ${\cal A}[Z',Z]$, ${\cal A}[Z',Z']$ excluding the anomalies $t_{ijk}$, the couplings $g_i$'s and charges $q^{f-A_i}$'s, and splitting them in the Axionic and GCS, the mass-independent and the mass-dependent contributions (for the fermion in the loop to be the electron or the muon $f=e/\m$).}\label{Fig:figCall}
\end{center}
\end{figure}

\subsection{Comparing contributions}\label{Comparing contributions}

In this section, we compare the   contribution
of the Axionic and GCS, the mass-independent and the mass-dependent contributions to the $g-2$ of the muon, for each of the diagrams  ${\cal A}[Z',\g]$, ${\cal A}[Z',Z]$, ${\cal A}[Z',Z']$.
We focus on the absolute value of the contribution excluding the couplings $g_i$'s and charges $q^{f-A_i}$'s and leaving the integrals.
We present our results in the plots \ref{Fig:figCall}.
\bi
\item The green line is the axionic and GCS contributions.

\item The orange line is the contribution of the mass-independent part of the triangle.

\item The blue/red lines are the contribution of the mass-dependent part of the triangle for circulating the electron/muon.

In our analysis, we do not allow the electron to couple to the $Z'$, therefore, this plot is irrelevant for our phenomenological analysis, however, we present it as a comparison to the rest of the plots in order to signify its importance.

\ei
Also, we evaluate the same coupling and charge independent parts for the diagrams where only SM fields are involved.
\bea
\ba{lllllllllllllll}
\textrm{Amplitude} &~~~&&~~~& \textrm{Axions\&GCS} &~~~&  \textrm{mass-ind} &~~~&  \textrm{mass-dep ($f=e$ or $\m$)}\\
{\cal A}[Z,\g]&&: &&
1.82\times 10^{-10} &&
 ~~5.27\times 10^{-12} && -1.92\times 10^{-11}\\
 {\cal A}[Z,Z]&&: &&
2.30\times 10^{-11} && -5.27\times 10^{-12} && -7.23\times 10^{-11}
\ea~~~~
\label{Contributions from only SM particles}
\eea
where, at this level of accuracy, the contribution of the mass-dependent part either for $f=e$ or $f=\m$ is the same.

Therefore, from the plots in \ref{Fig:figCall} and the results in \eqref{Contributions from only SM particles}, we conclude that the leading contribution to the $g-2$ is coming from the ${\cal A}[Z',\g]$ diagram and especially from the Axionic and GCS part.

Remember that these contributions are multiplied with all the couplings, charges, and anomalies in order to evaluate the full contribution to the $g-2$, presented in section \ref{the anomalous contribution}.



\providecommand{\href}[2]{#2}\begingroup\raggedright

\end{document}